\documentclass[traditabstract]{aa}
\usepackage{graphicx}
\usepackage{txfonts}
\usepackage{todonotes}
\usepackage[normalem]{ulem}
\usepackage{amstext}
\usepackage[breaklinks=true]{hyperref}
\usepackage{xspace}
\usepackage{lscape}

\usepackage{listings}
\makeatletter
\renewcommand*\maketitle{%
  \thispagestyle{firstpage}
\begingroup
    \if@wideboxfn
    \setlength\bibindent{1.4\parindent}
    \else
    \setlength\bibindent{\parindent}
    \fi
    \renewcommand*\thefootnote{\@fnsymbol\c@footnote}%
    \renewcommand\@makefntext[1]{%
    \ifaa@longfn\hsize\textwidth\fi
    \noindent
    \hb@xt@\bibindent{\hss\@makefnmark\enspace}##1}
  \ifaa@twocolumn
  \begingroup
    \begin{aa@strip}
          \aa@maketitle
    \end{aa@strip}
    \@thanks	  	
  \endgroup
  \else
    \begingroup
      \let\thanks\footnote
      \aa@maketitle
    \endgroup
  \fi
\endgroup
  \setcounter{footnote}{0}%
}
\makeatother

\definecolor{dkgreen}{rgb}{0,0.6,0}
\definecolor{gray}{rgb}{0.5,0.5,0.5}
\definecolor{mauve}{rgb}{0.58,0,0.82}
\newcommand{\orcit}[1]{\protect\href{https://orcid.org/#1}{\protect\includegraphics[width=8pt]{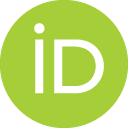}}}

\newcommand{\Gaia}{\textit{Gaia}\xspace}

\def\ltsima{$\, \buildrel < \over \sim \,$}
\def\simlt{\lower.5ex\hbox{\ltsima}}
\def\gtsima{$\, \buildrel > \over \sim \,$}
\def\simgt{\lower.5ex\hbox{\gtsima}}

\begin{document}

\title{Photometric metallicity for 694233 Galactic giant stars from \Gaia DR3 synthetic Str\"omgren photometry\thanks{Table 2 is available in its entirety only in electronic form at the CDS via anonymous ftp to \url{cdsarc.cds.unistra.fr} (130.79.128.5) or via
\url{https://cdsarc.cds.unistra.fr/cgi-bin/qcat?J/A+A/}}} 
\subtitle{Metallicity distribution functions of halo sub-structures}
\authorrunning{M. Bellazzini et al.}
\titlerunning{Photometric metallicity for 694233 giant stars}

\author{
M.~                    Bellazzini\orcit{0000-0001-8200-810X}\inst{1}
\and         D.~                       Massari\orcit{0000-0001-8892-4301}\inst{1}
\and         F.~                     De Angeli\orcit{0000-0003-1879-0488}\inst{2}
\and         A.~                     Mucciarelli\orcit{0000-0001-9158-8580}\inst{3,1}
\and         A.~                     Bragaglia\orcit{0000-0002-0338-7883}\inst{1}
\and         M.~                        Riello\orcit{0000-0002-3134-0935}\inst{2}
\and         P.~                   Montegriffo\orcit{0000-0001-5013-5948}\inst{1}
}
\institute{
INAF - Osservatorio di Astrofisica e Scienza dello Spazio di Bologna, via Piero Gobetti 93/3, 40129 Bologna, Italy%\relax                                                                                                                                                                                                                                      %\label{inst:0001}\vfill
\and 
Institute of Astronomy, University of Cambridge, Madingley Road, Cambridge CB3 0HA, United Kingdom%\relax                                                                                                                                                                                                                                                      %\label{inst:0002}\vfill
\and 
Dipartimento di Fisica e Astronomia, Universit\`a degli Studi di Bologna, Via Piero Gobetti 93/2, 40129 Bologna, Italy$\relax                                                                                                                                                                                                                                                      $\label{inst:0003}\vfill
}

%\date{Received date; Accepted date}

%\author{
%P. Montegriffo
%\inst{1}
%\and
%others\inst{2}\fnmsep\thanks{Just to show the usage of the elements in the author field}}

\date{Received on January 17, 2023; Accepted on April 17, 2023}

\abstract{We use the calibrations by Calamida et al. and by Hilker et al., and the standardised synthetic photometry in the 
$v$, $b$, and $y$ Str\"omgren passbands from \Gaia DR3 BP/RP spectra, to obtain photometric metallicities for a selected sample of 694233 old Galactic giant stars having |b|$>20.0\degr$ and parallax uncertainties lower than 10$\%$. 
The zero point of both sets of photometric metallicities has been shifted to
to ensure optimal match with the spectroscopic [Fe/H] values for 44785 stars in common with APOGEE DR17, focusing on the metallicity range where they provide the highest accuracy. The metallicities derived in this way from the Calamida et al. calibration
display a typical accuracy of $\simlt 0.1$~dex and 1$\sigma$ precision $\simlt 0.2$~dex in the range 
$-2.2 \simlt {\rm [Fe/H]}\simlt -0.4$, while 
they show a systematic trend with [Fe/H] at higher metallicity, beyond the applicability range of the relation. Those derived from the Hilker et al. calibration display, in general, worse precision, and lower accuracy in the metal-poor regime, but have a median accuracy $< 0.05$~dex for ${\rm [Fe/H]\ge -0.8}$. These results are confirmed and, consequently, the metallicities validated, by comparison with large sets of spectroscopic metallicities from various surveys.
The newly obtained metallicities are used to derive metallicity distributions for several previously identified sub-structures in the Galactic halo with an unprecedented number of stars. The catalogue including both sets of metallicities and the associated uncertainties is made publicly available.}


\maketitle

\section{Introduction}
\label{sec:introduction}

Photometric metallicities have a long and venerable history. For example, they were a key observational ingredient for both the most seminal papers on the formation of the Milky Way of the past century, i.e., \citet{els62} and \citet{sz78}. After a long period in which the technique was relegated to a minor role, it is gaining considerable credit and growing momentum in recent times \citep[see, among the most recent examples,][]{pristine_all,pristine_dwarfs,pristine_inn,skymap21,yang22,xu22,li22,lucey22,fallows22,an22,rix22,chandra22}.

In our view, there are two main factors driving this revival. First, the advent of modern digital surveys providing high precision photometry over wide areas of the sky \citep[like, e.g., 2MASS, SDSS, PS1, see][respectively, and references therein]{2mass,Thanjavur2021,Magnier2020}, some of which especially focused on photometric metallicities (as, for example, Pristine, \citealt{pristine_all} and SkyMapper, \citealt{skymap21}). Second, the availability of large samples of stellar abundances from modern spectroscopic surveys \citep[like, e.g., APOGEE, Gaia-ESO, GALAH, LAMOST, see][respectively]{apototal,ges_field,galah_dr3,lamost_tot}, allowing robust empirical calibration and validation of photometric metallicities. 

An additional factor facilitating the production and the successful scientific use of large samples of photometric metallicities is the advent of the various data releases of the ESA/\Gaia astrometric mission \citep[see][and references therein]{Prusti2016,ruwe,Gaia2021,dr3_general}. These greatly boosted our capability of unveiling the key connections between chemical and kinematical properties of stars and, for instance, pushed our ability to reliably distinguish giants from dwarfs to an unprecedented level, at least for stars with high precision parallaxes. It may be worth noting here that \Gaia is also beginning to provide a significant contribution in term of  chemical abundances from medium-resolution spectra in the Ca triplet region \citep{gspspec,dr3_chimica}, providing an additional, all sky, spectroscopic survey.

With \Gaia DR3 \citep{dr3_general}, the first release of spectrophotometry from the BP/RP (hereafter XP) low  resolution spectrometers\footnote{The Blue Photometer (BP) and the Red Photometer (RP) are the instruments on board of the Gaia spacecraft providing low resolution ($R\simeq 20-80$) spectro-photometry of astronomical sources, in the ranges $330~{\rm nm}\le \lambda\le 680~{\rm nm}$ and $640~{\rm nm}\le \lambda\le 1050~{\rm nm}$, respectively \citep{Prusti2016}. The external calibration process joins the two output spectra together into a single continuous flux table from 330~nm to 1050~nm \citep{dr3_ec_spectra}. For brevity, we refer to the spectra from the BP and RP instruments as well as the merged externally calibrated spectra as to ``XP spectra'', where X stands in place of both B and R in the XP acronym \citep{dpacp93}.}  \citep{Carrasco2021,dr3_spectra_fda,dr3_ec_spectra} opened a completely new window to our capability of deriving metallicities for a huge number of stars. \citet{dr3_gspphot} demonstrated the enormous potential of the full forward modelling of the entire XP spectra to derive stellar astrophysical parameters, that, however, will be fully exploited only in future \Gaia data releases, when several issues related to systematics in XP spectra and the tuning of their complex modelling will be fixed.

On the other hand, \citet{dpacp93} showed that very precise all-sky space-based synthetic photometry (SP) can be obtained from XP spectra for any medium/wide passband whose transmission curve (TC) is completely enclosed in the XP spectral range (330~nm-1100~nm). The residual systematics in externally calibrated XP spectra \citep{dr3_ec_spectra} still prevent accurate reproduction of existing photometry in some regions of the spectra, notably below $\simeq 550$~nm and, especially at $\lambda <400$~nm. However, using reliable sets of external photometric standards for a second-level calibration, with a process they denote as ``standardisation'', \citet{dpacp93} were able to reproduce existing photometry in several widely used systems  to millimag accuracy, over large colour ranges. In \citet{dpacp93} it was also demonstrated that narrow/medium band synthetic photometry is able to efficiently extract the information on metallicity, and, to a lesser extent, also on $\alpha$ element abundance, encoded in XP spectra \citep[see also][]{rix22,chandra22}. In fact, synthetic photometry with well suited passbands, on one side focuses on the region of the spectrum that is most sensitive to the the information of interest, and on the other maximises the signal to noise ratio (SNR) by making use of all the photons in the considered wavelength range. 

The Str\"omgren index $m_1 = (v-b)-(b-y)$ has been widely used to estimate metallicity of giant stars \citep[see, e.g.,][]{richter99, hilker00, att2000, dirsch00, faria07,cala_omega,arna10, piatti20, narloch21, narloch22}. Str\"omgren $vby$ is the only widely used medium-width band system for which XP SP has been standardised by
\citet{dpacp93}. Moreover, $m_1$ is the only photometric tracer of metallicity that does not imply the use of a passband sampling the critical region of XP spectra below 400~nm, among those explored by \citet{dpacp93}.

In this paper we follow up the successful experiments illustrated in \citet{dpacp93}, and we use standardised synthetic Str\"omgren $vby$ photometry from \Gaia DR3 XP spectra to obtain metallicity estimates for $\simeq 700000$ Galactic red giant branch (RGB), red horizontal branch (RHB), and asymptotic giant branch (AGB) stars fainter than the RGB tip, making the catalogue publicly available. As a first step, we limit to stars at relatively high Galactic latitude (|b|$>20.0\degr$) and having parallax uncertainties lower than 10\%, thus ensuring a proper separation between giants and dwarfs. The analysis and validation presented here provide the basis for future, more extensive applications.

It is very likely that the optimal way to obtain metallicity estimates from photometry and/or very low resolution spectrophotometry is by making use of as many indices as possible through extensively trained machine learning algorithms, as done, for instance, by \citet{fallows22,lucey22,yang22,chandra22}. \citet{rix22}, in particular, combined \Gaia XP spectra, photometric indices from XP SP and infrared AllWISE  photometry \citep{allwise} to get metallicities with typical precision of about 0.1~dex with an efficient machine learning algorithm trained with APOGEE metallicities. Here we show that, under certain conditions, we can achieve similar performances using a completely transparent analytic relation between Str\"omgren colour indices and [Fe/H], taking advantage of the excellent precision of standardised Str\"omgren XP SP from \Gaia DR3 and of the possibility to refine and validate the results using large samples of well-measured spectroscopic metallicities. 

We decided to rely on two existing and widely used calibrations, those by \citet[][C07 hereafter]{calamida07} and by 
\citet[][H00 hereafter]{hilker00}. Then we used a large sample of well-measured spectroscopic metallicities from APOGEE DR17 \citep{apogee_dr17} to adjust the zero point of the resulting metallicity scales to the APOGEE DR17 scale. This a posteriori re-calibration simultaneously accounts for differences in the adopted metallicity scale between the calibrating relations and APOGEE DR17 as well as for inhomogeneities in the calibration of Str\"omgren indices between the various authors and \Gaia XP synthetic photometry. The metallicities we derived from the C07 calibration are found to be more accurate and precise than those obtained from the H00 one in the range ${\rm [Fe/H]}\la-0.5$. Since we are more interested in the metal-poor regime we tailored our sample selection to the properties of the C07 relation (see Sect.~\ref{sample}).

The results shown here provide a simple recipe to derive reliable photometric metallicity for old giant stars from well measured Str\"omgren XP SP that can be of general use. 

The plan of the paper is the following. In Sect.~\ref{relation} we present the adopted calibrating relations and we briefly discuss the dependency of the photometric metallicity on age. In Sect.~\ref{sample} we present the adopted sample, with all the selections we applied. In Sect.~\ref{apogee} we show the comparison with APOGEE spectroscopic metallicity and the re-calibration of our photometric metallicities to the APOGEE DR17 scale. In Sect.~\ref{chemokin} we use our dataset to study the metallicity distribution function of several known substructures in the local Galactic halo, based on much larger samples than those previously considered in the literature. Finally, in Sect~\ref{conclu} we briefly summarise and discuss the main results of our analysis. In Appendix~\ref{appe_val}, we validate our photometric metallicities against their spectroscopic counterparts from various surveys. In Appendix~\ref{app_gctest} we use known member stars of star clusters included in our dataset to explore the performances of our photometric metallicities in different regimes. In Appendix~\ref{appe_ebv} and Appendix~\ref{appe_trend} we discuss the impact of reddening and of $\alpha$ elements abundance on our metallicity estimates, respectively.

\section{The calibrating relations}
\label{relation}

In the following, we will correct Str\"omgren indices from reddening as recommended by C07, i.e. adopting the extinction laws  $m_{1,0}=m_1+0.24E(B-V)$,  $(v-y)_0=(v-y)-1.24E(B-V)$, and $(b-y)_0=(b-y)-0.74E(B-V)$, from \citet{cm76}.

As our preferred choice, we convert reddening-corrected Str\"omgren indices $m_{1,0}$ and $(v-y)_0$ into metallicity estimates using the empirical relation:

\begin{equation}
\label{cal}
{\rm [Fe/H]_{phot,v}}=\frac{m_{1,0}-0.513(v-y)_0+0.312}{0.154(v-y)_0-0.096}
\end{equation}

\noindent
calibrated by C07 on selected Galactic globular clusters in the range $-2.2\le {\rm [Fe/H]}\le -0.7$, adopting the \citet{zw84} metallicity scale\footnote{In Appendix~\ref{app_semi} we briefly explore the performances of two additional calibrating relations provided by C07.}. According to C07 the uncertainty associated to the calibration itself is $\simeq 0.10$~dex.
C07 note that the relation is linear over the range $0.85\la (v-y)_0\la 3.0$. However, we preferred to limit, conservatively, to the narrower range $1.0< (v-y)_0<2.4$, also using their Fig.~11 as a guideline. Our choice was aimed (a) on the blue side, at excluding the colour range most prone to contamination from non-RGB stars (see Fig.~\ref{fig:cmd_vy0}) and where the sensitivity of $m_{1,0}$ to metallicity is at its minimum, and (b) on the red side, at avoiding the colour range that can be reached only by relatively metal-rich stars (see Fig.~\ref{fig:m0_apo}), thus minimising selection biases in metallicity distribution functions.

All the relations provided by C07 are calibrated on old stellar populations or old age stellar models and on a metallicity range never exceeding $-2.6\le {\rm [Fe/H]}\le -0.6$ (see Appendix~\ref{app_semi}). In principle they should not be used beyond their applicability range. Hence, their performance may not be optimal for young and/or very metal-rich or very metal-poor stars (see Sect.~\ref{apogee}, for further discussion).

Alternatively, we used the relation provided by H00, calibrated using globular clusters and field stars in the range $-2.2\le {\rm [Fe/H]}\le 0.0$:

\begin{equation}
\label{calH}
{\rm [Fe/H]_{phot,b}}=\frac{m_{1,0}-1.277(b-y)_0+0.331}{0.324(b-y)_0-0.032}
\end{equation}

\noindent
with metallicity in the \citet{zw84} scale. The colour range of validity is $0.5\le (b-y)_0\le 1.1$. The uncertainty associated to the calibration, as measured by H00, is in the range $0.11-0.16$~dex, after the rejection of stars whose absolute difference between [Fe/H]$_{phot}$ and [Fe/H]$_{spec}$ was larger than 0.25~dex.

The subscripts $v$ and $b$ assigned to the two different estimates of the photometric metallicity refer to the dependency on the $(v-y)_0$ and $(b-y)_0$ colours of the C07 and H00 calibrations, respectively.
1$\sigma$ errors on [Fe/H]$_{phot}$ from both relations have been computed following \citet{piatti20}, who provides error propagation formulae for the C07 and H00 calibrations, including the contribution of uncertainties on the coefficients of the calibrating relations. 
Uncertainties on observed magnitudes, computed as $\sigma_{\rm mag}=1.086\frac{\sigma Flux}{Flux}$, have been combined with uncertainties on extinction to compute uncertainties in the extinction-corrected indices and colours actually used in Eq.~\ref{cal} and Eq.~\ref{calH}. In practice, uncertainties on all the input quantities of these equations, including coefficients, are propagated into the uncertainty of the derived photometric metallicities.

The comparison with reliable spectroscopic samples, shown below, demonstrates that the typical standard deviation of the differences between photometric and spectroscopic metallicities is of similar order as the computed uncertainties. Still, these cannot properly account for all the sources of uncertainty, hence they must be interpreted with caution. Individual uncertainties should be considered as educated guesses, mainly driven by the signal-to-noise ratio of the magnitudes used to derive the Str\"omgren indices  and by the reddening  (see also Appendix~\ref{app_gctest}). 

\begin{figure*}[ht!]
\center{
\includegraphics[width=\columnwidth]{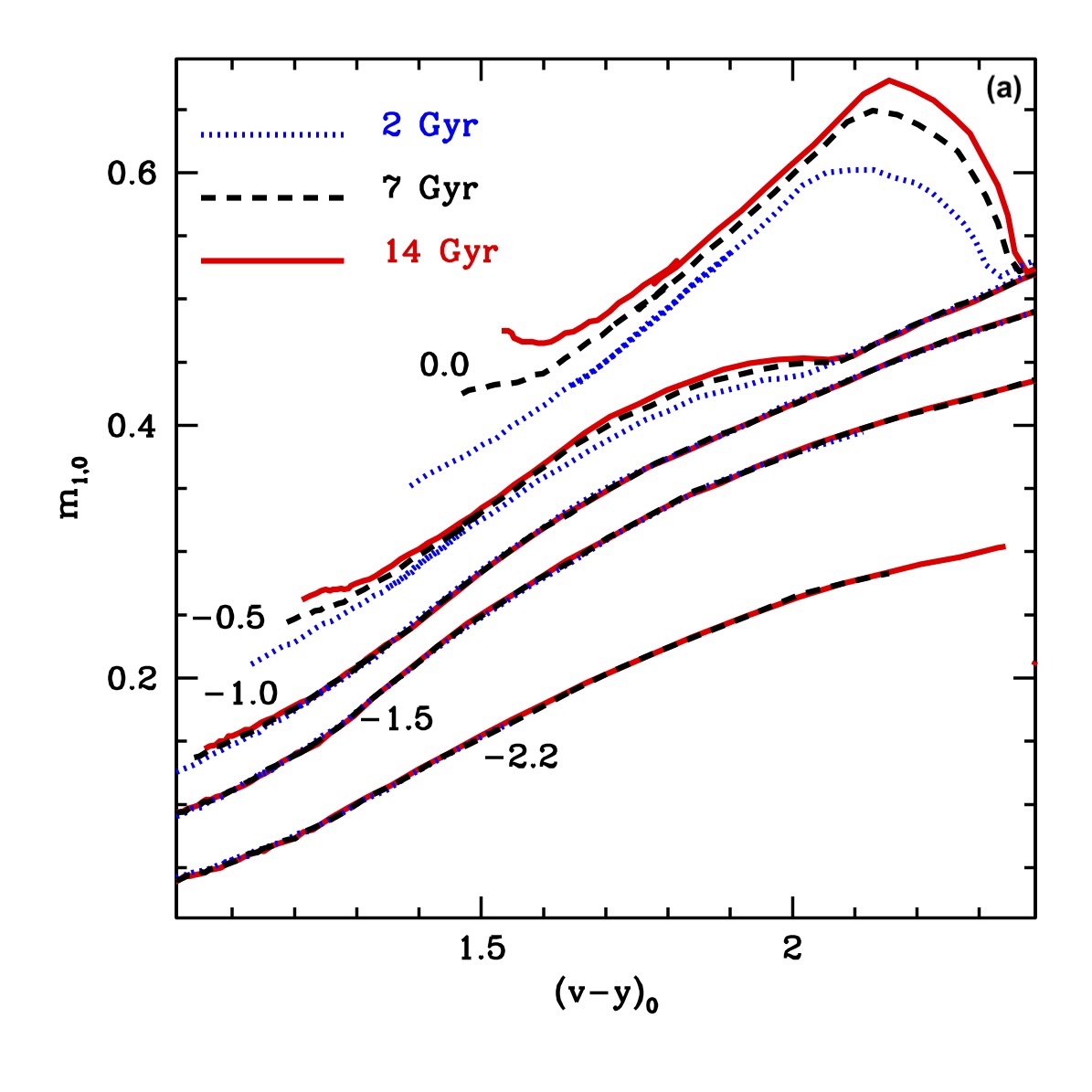}
\includegraphics[width=\columnwidth]{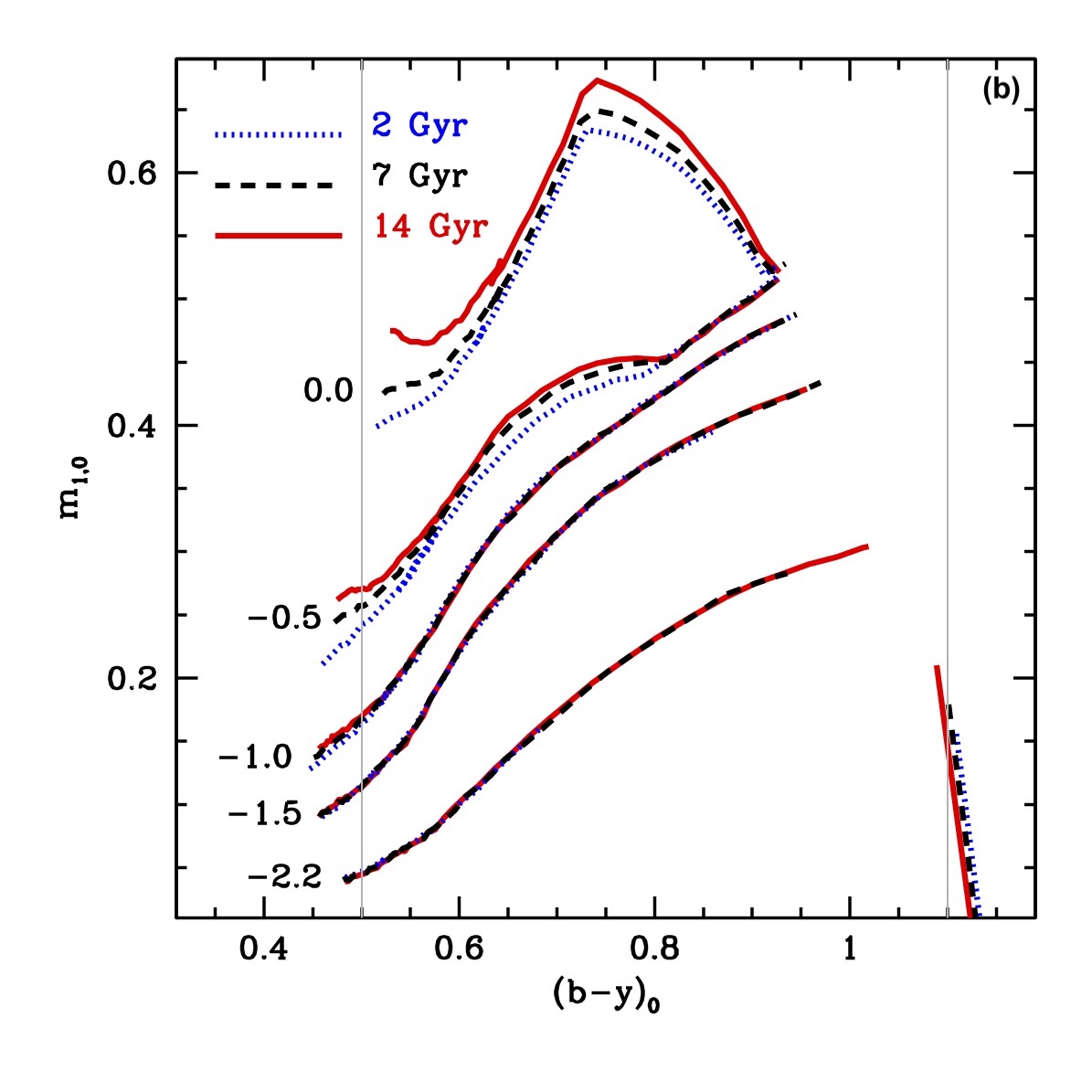}
}
\caption{RGB isochrones from the PARSEC set \citep{Bressan2012,Chen2019} in the $(v-y)_0$ -$m_{1,0}$ plane (left panel) and in the $(b-y)_0$ -$m_{1,0}$ plane (right panel). The five bundles of three isochrones of different ages (2, 7, and 14 Gyr) are labelled with their global metallicity ${\rm [M/H]}$=0.0, -0.5, -1.0, -1.5, -2.2, from top to bottom, respectively. The plot in the left panel is limited to the $(v-y)_0$ colour range in which we actually compute photometric metallicities. In the right panel we plotted only the part of the isochrones whose  $(v-y)_0$ colour is in the permitted range, while the range of validity of the H00 calibration in $(b-y)_0$ is enclosed between the two grey vertical lines.}
\label{fig:agemet}
\end{figure*} 

Metallicity derived from the $m_{1,0}$ index for RGB stars are generally believed to be nearly independent of age \citep[see, e.g.,][]{hilker00,dirsch00,faria07,narloch22}, but arguments against this view have been recently presented by \citet{piatti20} and \citet{dpacp93}. In Fig.~\ref{fig:agemet} we show the results of a small experiment we made using PARSEC solar-scaled isochrones \citep{Bressan2012,Chen2019} produced with the dedicated web tool\footnote{http://stev.oapd.inaf.it/cgi-bin/cmd}.
Five bundles of RGB isochrones spanning the global metallicity (${\rm [M/H]}={\rm log(Z/X)-log(Z/X)_{\sun}}$) range from solar to $\simeq 0.006$ times solar are plotted in the  
$(v-y)_0$, $m_{1,0}$ plane (left panel) and in the $(b-y)_0$, $m_{1,0}$ plane (right panel). Each bundle is composed by three isochrones of the same metallicity but having age 2, 7, and 14 Gyr. The figure strongly support the hypothesis of age independency of $m_{1,0}$, at least in the metal-poor regime. For ${\rm [M/H]}\le-1.0$ the isochrones of each bundle are virtually indistinguishable over the entire colour range of interest. At ${\rm [M/H]}=-0.5$ the differences between the 14~Gyr and the 7~Gyr isochrones are negligible, those with the 2~Gyr isochrone become relatively sizeable only in certain ranges of colour. It is only with the isochrone bundle at solar metallicity that age-metallicity degeneracies may be relevant, especially for $(v-y)_0\simgt 2.0$ ($(b-y)_0\simgt 0.7$). In this metallicity regime, the dependency on age is significantly weaker in the $(b-y)_0$, $m_{1,0}$ plane than in the $(v-y)_0$, $m_{1,0}$ plane.

The possible effect, as illustrated also by \citet{piatti20} and \citet{dpacp93}, is to underestimate the metallicity of stars significantly younger than the typical globular cluster. This behaviour should be kept in mind when interpreting results based on this kind of photometric metallicities, especially when the application of the C07 relation is extrapolated beyond the metal-rich limit of the calibration.

Another well known problem that may affect metallicities derived from $m_{1,0}$ is due to the fact that strong CN molecular features in the $v$ band can mimic the effect of higher metallicity \citep[see, e.g.][]{richter99, hilker00,narloch22}. There is no way to prevent or mitigate this issue, that, however, is not expected to affect a large number of field stars. The highest frequency of CN strong stars is observed to occur in old clusters \citep[see][for a brief discussion and references]{narloch22}. In the massive old globular cluster $\omega$~Cen \citet{cala_omega} estimates that 19\% $\pm$ 1\% of RGB stars are CN strong. The comparison between photometric and spectroscopic metallicities in Sect.~\ref{apogee} and in Appendix~\ref{appe_val} suggests that the fraction of CN strong stars is significantly lower than this in our sample of field stars (see Sect.~\ref{apogee} for discussion). In any case, according to these comparisons, the impact on our metallicity estimates is smaller than the typical uncertainty for the vast majority of the considered stars.

\section{The sample}
\label{sample}

We used the GaiaXPy tool\footnote{\url{https://Gaia-dpci.github.io/GaiaXPy-website/}} to get standardised $vby$ photometry from the \Gaia archive\footnote{\url{https://gea.esac.esa.int/archive/}}. We retrieved from the archive all the stars that (a) are included in the \Gaia Synthetic Photometry Catalogue \citep[GSPC,][]{dpacp93}, (b) have Galactic latitude |b|$>20.0\degr$, (c) {\tt parallax\_over\_error > 10}, (d) {\tt RUWE < 1.3} \citep{ruwe}, and (e) $|C^{\star}|<\sigma_{C^{\star}}(G)$, according to Eq.~18 of \citet{Riello2021}.

Condition (a) merely ensures that the stars have XP spectra from which at least one of the wide band magnitudes included in the GSPC can be obtained with SNR>30, while condition (b) is intended to avoid the most crowded and extinct regions of the Galaxy. Condition (c) guarantees that reliable geometric distances of all the stars can be obtained simply as the inverse of the parallax \citep{bj15}, thus allowing a simple and robust
selection of giant stars from the distance and reddening corrected Color Magnitude Diagram (CMD), while condition (d) and (e) are intended to remove from the sample stars with less reliable astrometry \citep{ruwe} and stars whose fluxes are contaminated by crowding and/or poor background subtraction \citep{Riello2021}. Finally, to include only stars with precisely measured Str\"omgren magnitudes we imposed that they must have SNR>20 in all the three $vby$ bands, that is {\tt StromgrenStd\_flux\_i/StromgrenStd\_flux\_error\_i>20} for {\tt i=v,b,y}. 

\begin{figure*}[!ht]
\center{
\includegraphics[scale=0.30]{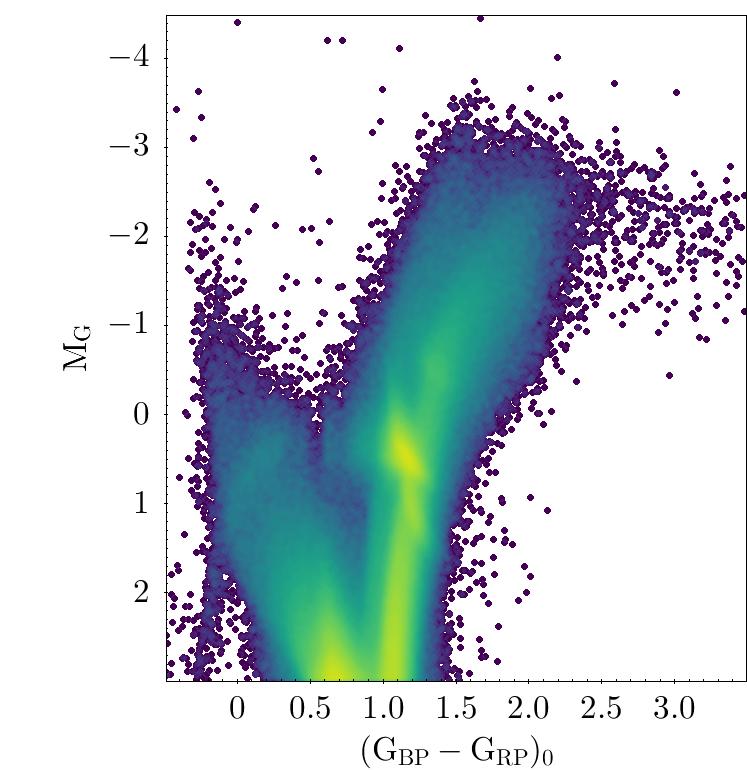}
\includegraphics[scale=0.30]{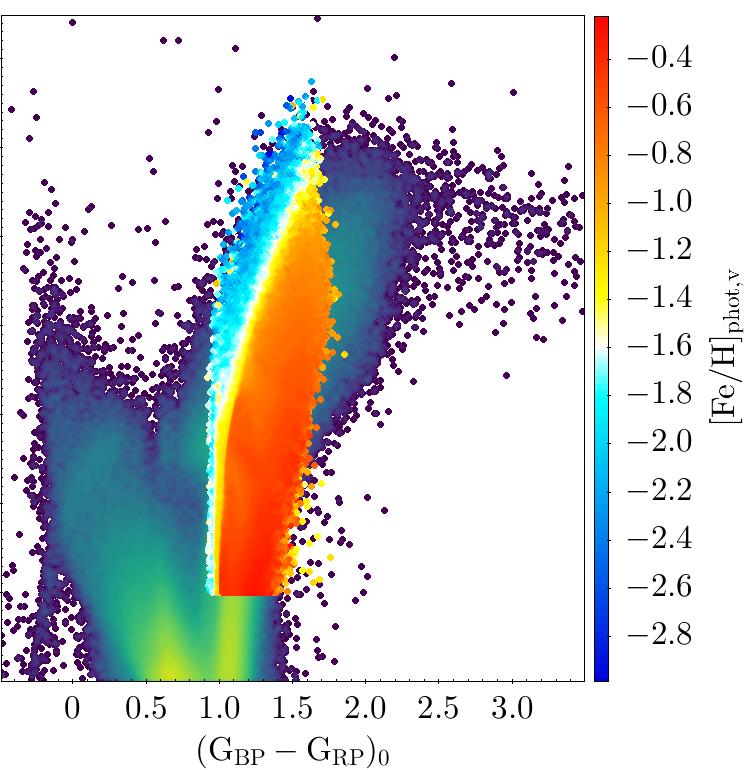}
}
\caption{Colour and magnitude distribution of selected stars. Panel (a): CMD in the \Gaia DR3 photometric system for the entire sample of 16,435,526 stars we extracted from the \Gaia archive. Stars are colour coded according to the logarithm of the local density. 
Panel (b): On the same CMD we superpose that of our final selected sample of old giants, with stars colour coded according to the mean local value of ${\rm [Fe/H]_{phot,v}}$.
\label{fig:cmd_bprp}
}
\end{figure*}

All these conditions lead to a sample of 16,435,526 stars, for which we obtained the reddening $E(B-V)$ by interpolating into the \citet{sfd98} maps, recalibrated according to \citet[][some discussion on this choice is provided in Appendix~\ref{appe_ebv}]{Shlafly2011}. A reddening uncertainty of $0.1{\rm E(B-V)}$ has been assumed in computing the uncertainty on photometric metallicity \citep{sfd98}. Then, we applied a first coarse selection of candidate RGB stars, keeping only those having $0.8<(G_{BP}-G_{RP})_0<2.2$ and
$-4.0<M_G<2.0$, hence reducing the sample to less than 900,000 stars. The $M_G<2.0$ cut is mainly aimed at limiting the contamination by non-RGB stars on the blue side of the selected sample, as moving toward fainter $M_G$ the RGB and the blue plume of young main sequence, sub giant and blue loop stars converge to similar colours, leading to some overlap between the two populations in the CMD (see Fig.~\ref{fig:cmd_bprp}). 
We also removed from the sample 11896 sources classified as non single, and 5615 sources classified as variables in the {\tt Gaiadr3.Gaia\_source} table\footnote{By keeping only stars with {\tt non\_single\_star==0} and {\tt phot\_variable\_flag!="VARIABLE"}.}, plus 23265 sources having $E(B-V)\ge 0.3$, to exclude stars whose Str\"omgren indices are significantly impacted by interstellar extinction. 

Limiting the $(v-y)_0$ range to the adopted validity range of Eq.~\ref{cal} removed 36477 additional stars. At this stage we found that 925 out of the 760535 stars in our sample had photometric metallicities outside the broad range $-5.0\le {\rm [Fe/H]_{phot,v}}\le +0.5$, and we excluded this handful of unrealistic estimates.

Finally, the inspection of the CMD with stars coloured according to their ${\rm [Fe/H]_{phot,v}}$ value, revealed the residual presence of stars whose location in the RGB was not consistent with their metallicity. Anomalously metal-rich stars in the blue side of the RGB were likely largely contributed by non-genuine RGB stars, i.e., mainly young He-burning stars, or, possibly, by some CN-strong RGBs. Anomalously metal-poor stars on the red side of the RGB are more difficult to interpret, still they turn out to constitute only about $0.1\%$ of the sample. Both class of stars were removed using metallicity-dependent conditions, similar to those adopted by \citet{hasse21} to avoid contamination from young stars in their sample of RGB stars in the Large Magellanic Clouds.
In practice, on the blue side we removed the stars having ${\rm [Fe/H]_{phot,v}}\ge -1.0$ whose colour was 
$(G_{BP}-G_{RP})_0<t(M_G)$  as well as those having $(G_{BP}-G_{RP})_0<0.98$ and ${\rm [Fe/H]_{phot,v}}\ge -1.5$ (61180 stars and 3160 stars, respectively); on the red side 
those having $(G_{BP}-G_{RP})_0>t(M_G)+0.1$ and 
${\rm [Fe/H]_{phot,v}}\le -1.5$ (772 stars),
where 
\begin{equation}
\label{t_G}
 t(M_G)=1.1 -0.1M_G +0.03M_G^2   
\end{equation}

\noindent
is a convenient threshold in colour approximately following the curvature of the RGB in the CMD.  
Finally we removed 265 stars with $\epsilon{\rm [Fe/H]_{phot,v}}\ge 0.4$~dex.

%###################################################################
%###################################################################

\begin{figure}[ht!]
\center{
\includegraphics[width=\columnwidth]{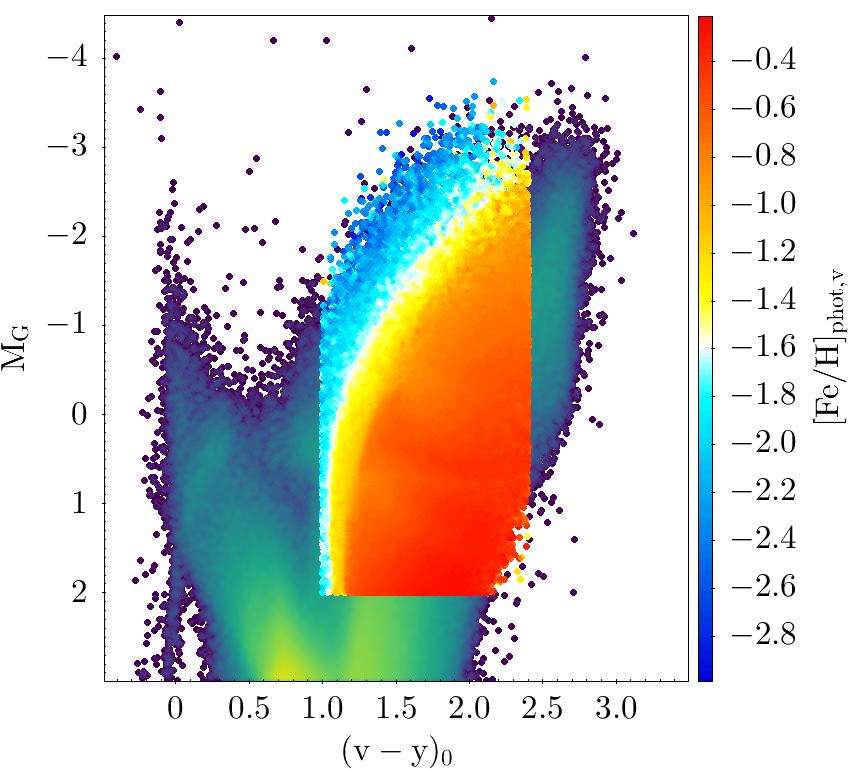}
}
\caption{The same as Fig.~\ref{fig:cmd_bprp}b but using $(v-y)_0$ colour, instead of $(G_{BP}-G_{RP})_0$}.
\label{fig:cmd_vy0}
\end{figure}

The main characteristics of the final sample, that contains 694233 stars, are illustrated in Fig.~\ref{fig:cmd_bprp}. In particular, Fig.~\ref{fig:cmd_bprp}a shows the distance and reddening corrected CMD of the entire sample of 16,435,526 stars that we extracted from the \Gaia archive, with stars colour coded according to the local density on the CMD, to highlight all the relevant evolutionary features of the diagram. In Fig.~\ref{fig:cmd_bprp}b, the stars belonging to the final sample are superposed to the the same overall CMD but colour coded according to the mean local value of ${\rm [Fe/H]_{phot,v}}$, illustrating both the colour-magnitude limits of the sample and the expected correlation between ${\rm [Fe/H]_{phot,v}}$  and the colour along the RGB. Fig.~\ref{fig:cmd_vy0} shows how clean and clear is this correlation if the $(v-y)_0$ colour is used, instead of $(G_{BP}-G_{RP})_0$, and illustrates very clearly the adopted thresholds in $(v-y)_0$.

The comparison between the two panels of Fig.~\ref{fig:cmd_bprp} shows also that, while the sample is dominated by RGB stars, a significant population of He-burning Red Clump (RC) stars, around $M_G\simeq +0.5$, is also included, as well as a minor contribution from AGB stars fainter than the RGB tip, the AGB clump of the metal-rich population being barely visible at $M_G\sim -0.4$ and $(G_{BP}-G_{RP})_0\sim 1.3$. Since $m_{1.0}$ is supposed to trace blanketing from metal lines in the range $\simeq 400-425$~nm in a gravity-independent way \citep[for giants;][]{arna10}, it should do it on stars of similar spectral type independently of their evolutionary stage, as, in any case, we are dealing with old stars (age$\simgt 2$~Gyr). A close inspection of Fig.~\ref{fig:cmd_bprp}b and Fig.~\ref{fig:cmd_vy0} reveals that, on average, RC stars are recognised to be more metal-rich than RGBs of the same colour, supporting this view. Indeed, the comparisons with spectroscopic metallicities (Sect.~\ref{apogee} and Appendix~\ref{appe_val}) demonstrate that reliable photometric metallicities can be obtained for all kind of stars included in our sample. We carefully verified that our photometric metallicities of RC and AGB stars are indistinguishable from those of RGB stars, when compared to their spectroscopic counterparts (see also Appendix~\ref{app_gctest}). 

It is clear that the assembled sample cannot be considered in any way complete, nor sampling the underlying metallicity distribution function (MDF) in a fully uniform/fair way, as it is subject to biases induced by the adopted colour- and, to a lesser extent, metallicity-dependent, selection criteria, in addition to those associated to the selections in latitude and extinction. Our main goal was the reliability of the metallicity estimates, which comes at the cost of excluding all the stars that we can identify as possibly having a badly measured metallicity.
However any bias is expected to have only mild effects on MDFs, especially in the metallicity range far from the extremes reached by our sample, where colour cuts and contamination by non-RGB stars may play a role, e.g, for ${\rm [Fe/H]}\la -2.2$ and ${\rm [Fe/H]}\ga 0.0$.

%\begin{figure*}
\begin{figure}
\center{
\includegraphics[width=\columnwidth]{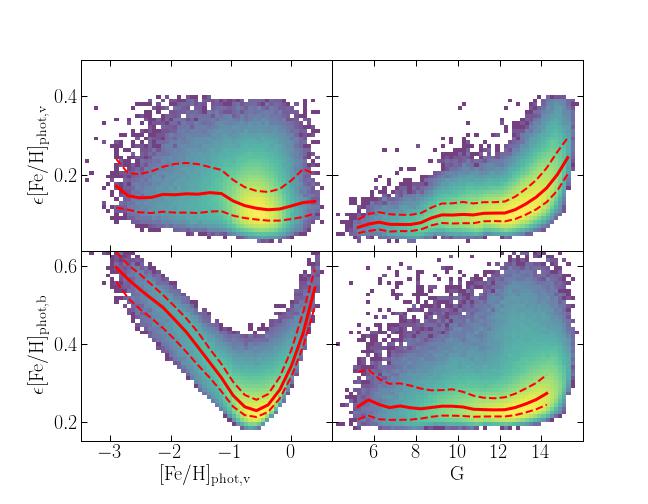}
}
\caption{Distributions of individual uncertainties on  ${\rm [Fe/H]_{phot,v}}$ (upper panels) and on ${\rm [Fe/H]_{phot,b}}$ (lower panels) as a function of ${\rm [Fe/H]_{phot,v}}$  (left panels) and of G magnitude (right panels). In all cases the colour scale is coded according to the logarithm of local density. In all panels, the continuous red curve traces the median of the $\epsilon {\rm [Fe/H]_{phot,v}}$ distribution, while the red dashed curves trace the 16th and 84th percentiles of the same distribution. Please note the different scales of the y-axis between the upper and lower panels}.
\label{fig:efe}
\end{figure} 
%\end{figure*} 

%%%%%%%%%%%%%%%%%%%
\begin{figure*}[ht!]
\center{
\includegraphics[width=\columnwidth]{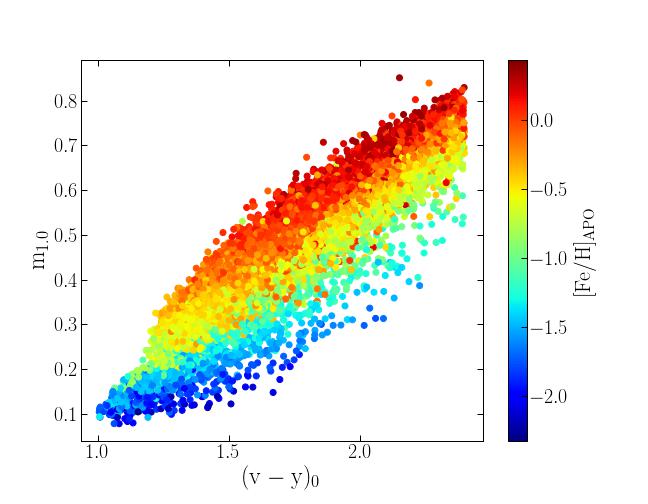}
\includegraphics[width=\columnwidth]{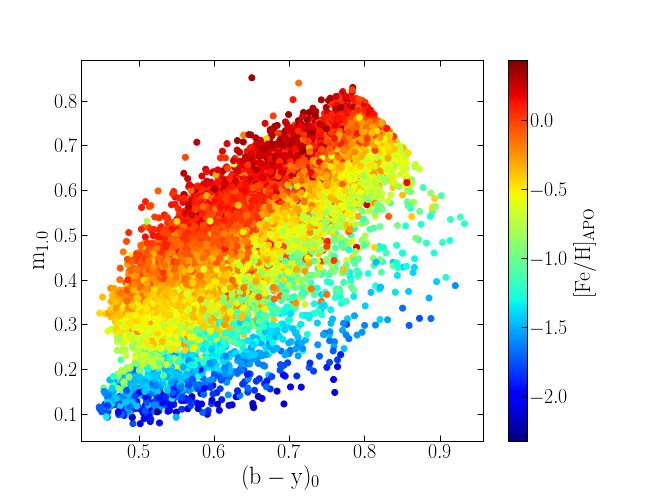}
}
\caption{Stars of the APOGEE VS plotted in the $(v-y)_0$, $m_{1,0}$ (left panel) and $(b-y)_0$, $m_{1,0}$ diagrams. Points are coloured according to their ${\rm [Fe/H]_{spec}}$ values from APOGEE DR17.}
\label{fig:m0_apo}
\end{figure*} 
%%%%%%%%%%%%%%%%%%%%%%%%%%%%%%%%%%%%%%%%%

Figure~\ref{fig:efe} displays the distribution of $\epsilon {\rm [Fe/H]_{phot,v}}$  and $\epsilon {\rm [Fe/H]_{phot,b}}$ as a function of G magnitude and ${\rm [Fe/H]_{phot,v}}$.  All stars in our sample have $G<16.0$, 50\% of them have $G\le 12.76$, and 95\% have $G\le 14.48$. The sample is dominated by metal rich stars, 50\% having ${\rm [Fe/H]_{phot,v}}\ge -0.41$ and 95\% having ${\rm [Fe/H]_{phot,v}}\ge -0.96$. $\epsilon {\rm [Fe/H]_{phot,v}}$ seems well behaved, with very mild dependency on ${\rm [Fe/H]_{phot,v}}$ and increasing median error at fainter magnitudes, as expected.
The median ($P_{50}$) uncertainties are $\simlt 0.15$~dex for nearly the entire ranges of G and ${\rm [Fe/H]_{phot,v}}$. The 99th percentile ($P_{99}$) is $<0.2$~dex for $G\simlt 12.5$. 
638551 of the 694233 stars included in our final sample have individual metallicity uncertainties $\epsilon {\rm [Fe/H]_{phot,v}}\le 0.2$~dex. 

On the other hand, $\epsilon {\rm [Fe/H]_{phot,b}}$ are significantly larger than $\epsilon {\rm [Fe/H]_{phot,v}}$, with a median as a function of magnitude of about 0.25~dex and strong trends with ${\rm [Fe/H]_{phot,v}}$. We  attribute this behaviour to the stronger dependency on colour of Eq.~\ref{calH} with respect to Eq.~\ref{cal} and to the significantly larger uncertainties associated to the coefficients of Eq.~\ref{calH} than to those of Eq.~\ref{cal}.

%%%%%%%%%%%%%%%%%%%
\begin{figure}[ht!]
\center{
\includegraphics[width=\columnwidth]{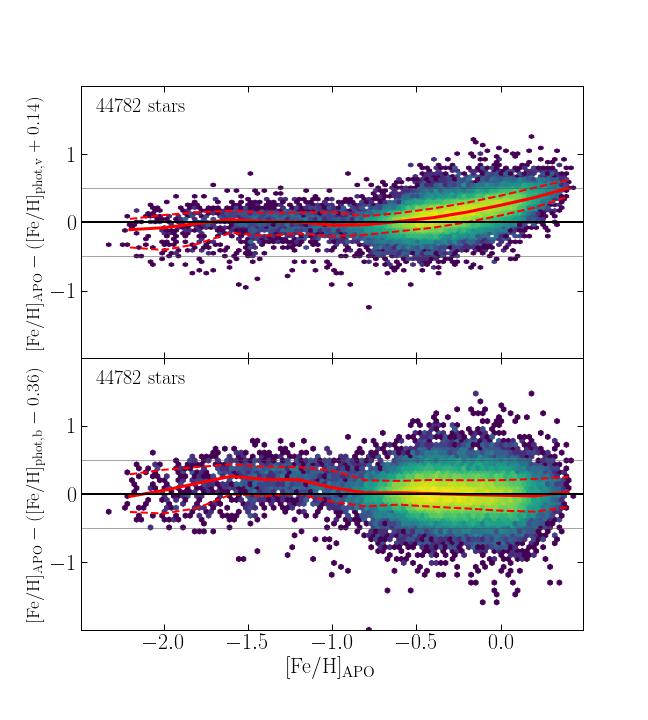}
}
\caption{The differences between the spectroscopic and photometric metallicities are plotted, as a function of spectroscopic metallicity, for the APOGEE VS, for ${\rm [Fe/H]_{phot,v}}$ (upper panel) and ${\rm [Fe/H]_{phot,b}}$
(lower panel). 
%The distributions are represented as 2-d histograms with hexagonal bins color-coded according to the logarithm of the local density. 
The horizontal grey lines are located at $\Delta {\rm [Fe/H]}=\pm 0.5$~dex. The continuous red lines trace $P_{50}$ and the dashed red lines $P_{16}$ and $P_{84}$ of the $\Delta {\rm [Fe/H]}$ distribution. A shift of $+0.14$~dex has been applied to ${\rm [Fe/H]_{phot,v}}$ to minimise the median $\Delta {\rm [Fe/H]}$ in the region ${\rm [Fe/H]_{spec}}\le -0.8$, while a shift of $-0.36$~dex has been applied to ${\rm [Fe/H]_{phot,b}}$ to minimise the median $\Delta {\rm [Fe/H]}$ in the region ${\rm [Fe/H]_{spec}}\ge -0.8$.}
\label{fig:apo_fe}
\end{figure} 
%%%%%%%%%%%%%%%%%%%%%%%%%%%%%%%%%%%%%%%%%

\section{Refining the calibration with APOGEE data}
\label{apogee}

Among the spectroscopic datasets we considered to validate our photometric metallicities (see Appendix~\ref{appe_val}), APOGEE DR17 \citep{apogee_dr17} has both a large number of well measured stars in common with our sample and a good sampling of stars in the metal-poor regime. 
Moreover, it is widely used to calibrate machine learning algorithms aimed at deriving photometric metallicities \citep[see, e.g.,][]{rix22,yang22} and for validation. For these reasons we adopt a selected set of APOGEE DR17 stars for a first validation of our dataset and to refine the calibration of ${\rm [Fe/H]_{phot}}$ by minimisation of the median difference between spectroscopic and photometric metallicities, always expressed as $\Delta {\rm [Fe/H]} = {\rm [Fe/H]_{spec}} - {\rm [Fe/H]_{phot}}$, in the following. 

%%%%%%%%%%%%%%%%%%%%%%%%%%%%%%%%%%%%%%%%%%%%%%%%%%%%%%%%%%%%
\begin{table*}[!htbp]
\centering
\caption{\label{tab:APOall_P} Comparison with the APOGEE VS. Percentiles 
of the $\Delta {\rm [Fe/H]_{C}}$ and $\Delta {\rm [Fe/H]_{H}}$ distributions. }
{
    \begin{tabular}{lccccccccr}
[Fe/H] &  $P_{50}^{[Fe/H]_C}$  &  $P_{16}^{[Fe/H]_C}$  &   $P_{84}^{[Fe/H]_C}$  & $\sigma_{[Fe/H]_C}$  &  $P_{50}^{[Fe/H]_H}$  &  $P_{16}^{[Fe/H]_H}$  &   $P_{84}^{[Fe/H]_H}$  & $\sigma_{[Fe/H]_H}$ &  n  \\  
\hline  
-2.20 & -0.11 & -0.37 &   0.05 &  0.21 & -0.03 & -0.26 &   0.29 &  0.28 &  17 	  \\
-2.00 & -0.08 & -0.41 &   0.11 &  0.26 &  0.05 & -0.28 &   0.36 &  0.32 &  70 	  \\
-1.80 & -0.02 & -0.32 &   0.15 &  0.23 &  0.16 & -0.21 &   0.40 &  0.30 &  121    \\
-1.60 &  0.04 & -0.15 &   0.17 &  0.16 &  0.26 & -0.00 &   0.44 &  0.22 &  182    \\
-1.40 &  0.01 & -0.20 &   0.14 &  0.17 &  0.21 & -0.03 &   0.40 &  0.21 &  327    \\
-1.20 &  0.01 & -0.16 &   0.14 &  0.15 &  0.21 &  0.00 &   0.40 &  0.20 &  339    \\
-1.00 & -0.04 & -0.21 &   0.15 &  0.18 &  0.10 & -0.11 &   0.34 &  0.23 &  372    \\
-0.80 & -0.03 & -0.18 &   0.09 &  0.14 &  0.02 & -0.18 &   0.20 &  0.19 &  1258   \\
-0.60 &  0.01 & -0.13 &   0.14 &  0.13 &  0.02 & -0.15 &   0.19 &  0.17 &  6397   \\
-0.40 &  0.07 & -0.08 &   0.20 &  0.14 &  0.01 & -0.19 &   0.21 &  0.20 &  13001  \\
-0.20 &  0.15 &  0.00 &   0.29 &  0.14 & -0.01 & -0.21 &   0.20 &  0.21 &  11551  \\
 ~0.00 &  0.25 &  0.10 &   0.39 &  0.14 & -0.02 & -0.24 &   0.20 &  0.22 &  8424   \\
 ~0.20 &  0.36 &  0.21 &   0.51 &  0.15 & -0.03 & -0.26 &   0.22 &  0.24 &  2519   \\
 ~0.40 &  0.50 &  0.38 &   0.62 &  0.12 &  0.03 & -0.19 &   0.25 &  0.22  &  203    \\
\hline
    \end{tabular}
}
\tablefoot{ ${\rm [Fe/H]_{C}}$ and ${\rm [Fe/H]_{H}}$ are defined in Eq.~\ref{eq:fe} and Eq.~\ref{eq:fe_h}, respectively. $\sigma$ are defined as 
$\sigma=0.5(P_{84}-P_{16})$.  
n is the number of sources in the considered bin.}
\end{table*}
%%%%%%%%%%%%%%%%%%%%%%%%%%%%
%%%%%%%%%%%%%%%%%%%%%%%%%%%%%%%%%%%%%%%%%%%%%%%%%%%%%%%%%%%% 

To select a set of well measured APOGEE DR17 metallicities to compare with, we retain only stars having 
{\tt SNR$\ge$ 70}, {\tt ASPCAPFLAG=0},  {\tt FE\_H\_ERR<0.1}, and  {\tt ALPHA\_M\_ERR<0.2}
\footnote{See \url{https://data.sdss.org/datamodel/files/APOGEE_ASPCAP/APRED_VERS/ASPCAP_VERS/allStarLite.html} for definitions.}. We found 44782 of these stars in common with our sample, that hereafter we denote as the APOGEE Validating Sample (VS). Figure~\ref{fig:m0_apo} illustrates very clearly how well the $m_{1,0}$ index we derived from \Gaia XP synthetic photometry correlates with spectroscopic metallicities from the APOGEE VS, for a given $(v-y)_0$ or $(b-y)_0$ colour.

The distributions  of $\Delta {\rm [Fe/H]}$ as a function of ${\rm [Fe/H]_{spec}}$ for 
${\rm [Fe/H]_{phot,v}}$ and ${\rm [Fe/H]_{phot,b}}$
are shown in Fig.~\ref{fig:apo_fe}. In this figure, as well as in all the other figures of this kind in the following, the distributions are represented as 2-d histograms, with hexagonal bins colour-coded according to the logarithm of the number of stars in the bin. The continuous red line traces the median ($P_{50}$) of the $\Delta {\rm [Fe/H]}$ distribution computed in bins of 0.2~dex width, while the dashed red lines enclose the 15.87th ($P_{16}$) and the 84.13th ($P_{84}$) percentiles, as a proxy for the $\pm 1\sigma$ interval around the median.

%%%%%%%%%%%%%%%%%%%%%%%%%%%%%%%%%%%%%%%%%%%%%%%%%%%%%%%%%%%%
\begin{table*}[!htbp]
\centering
\caption{\label{tab:sample} First five rows of our dataset of photometric metallicities.}
{
    \begin{tabular}{lccccc}
 \Gaia DR3 source\_id   & ${\rm[Fe/H]_{C}}$ & ${\rm\epsilon[Fe/H]_{C}}$ &  ${\rm[Fe/H]_{H}}$     &   ${\rm\epsilon[Fe/H]_{H}}$  & {\tt flagH}   \\
\hline
   4442257706155964416 & -0.89 &  0.18 & -1.02 & 0.30 & 0 \\
   2159113251207531008 & -0.38 &  0.10 & -0.27 & 0.22 & 0 \\
    847127671447383168 & -0.35 &  0.07 & -0.26 & 0.20 & 0 \\
    852805961810658432 & -0.55 &  0.12 & -0.53 & 0.23 & 0 \\
   4581710552614214144 & -0.62 &  0.08 & -0.53 & 0.20 & 0 \\
        continues & ... & ... & ... & ...& ... \\
\hline
    \end{tabular}
}
\tablefoot{flagH=1 indicates stars whose $(b-y)_0$ color is slightly beyond the validity limit of the H00 calibration, on the blue side.}
\end{table*}
%%%%%%%%%%%%%%%%%%%%%%%%%%%%%%%%%%%%%%%%%%%%%%%%%%%%%%%%%%%% 

A $+0.14$~dex shift has been applied to ${\rm [Fe/H]_{phot,v}}$, to minimise the median $\Delta {\rm [Fe/H]}$ in the region ${\rm [Fe/H]_{spec}}\le -0.7$, corresponding to the validity range of the C07 calibration, where there is no perceivable trend  between $\Delta {\rm [Fe/H]}$ and  ${\rm [Fe/H]_{spec}}$. Once this shift is applied the spectroscopic metallicities are reproduced with a median accuracy of $\le 0.10$ dex over the range $-2.0\le {\rm [Fe/H]_{spec}}\le -0.4$, while a trend of $\Delta {\rm [Fe/H]}$ with ${\rm [Fe/H]_{spec}}$ arises at higher metallicities, reaching a median  offset of +0.50~dex at ${\rm [Fe/H]_{spec}}=+0.4$ (see Table~\ref{tab:APOall_P}). The sense of the trend is that spectroscopic metallicities are underestimated by their photometric counterparts for ${\rm [Fe/H]_{spec}}\ga -0.4$. It is important to note here that this trend with metallicity arises in a metallicity regime that is well beyond the strict applicability range of the C07 calibration ($-2.2\le {\rm [Fe/H]}\le -0.7$), hence it does not come as a surprise. 
For the entire sample, the semi-difference between $P_{84}$ and $P_{16}$, that we take here as a proxy for the standard deviation ($\sigma$ hereafter), is $< 0.20$~dex for  ${\rm [Fe/H]_{spec}}\ge -1.6$ and $< 0.15$~dex for  ${\rm [Fe/H]_{spec}}\ge -0.8$, ranging between 0.21~dex and 0.26~dex in the most metal-poor, sparsely populated bins,  at ${\rm [Fe/H]_{spec}}\le -1.8$. If not otherwise stated we will denote this quantity, that traces the {\em precision} of our measures, as $\sigma$. 

A $-0.36$~dex shift has been applied to ${\rm [Fe/H]_{phot,b}}$, to minimise the median $\Delta {\rm [Fe/H]}$ in the region ${\rm [Fe/H]_{spec}}\ge -0.8$, where there is virtually no trend  between $\Delta {\rm [Fe/H]}$ and  ${\rm [Fe/H]_{spec}}$. Once this shift is applied the spectroscopic metallicities are reproduced with a median accuracy of $\le 0.04$ dex over the range $-0.8\le {\rm [Fe/H]_{spec}}\le +0.4$, while at lower metallicity, the deviations of $\Delta {\rm [Fe/H]}$ reach amplitudes $\ga 0.2$~dex. $\sigma$ ranges between 0.17~dex and 0.32~dex, and it is $\ge 0.2$~dex in all bins except two (Tab.~\ref{tab:APOall_P}).

The features emerging from the comparison of our photometric metallicities with the APOGEE DR17 VS and described above are found also when the comparison is performed with validating samples from other spectroscopic surveys, as shown in Appendix~\ref{appe_val}.
The overall performances of our corrected ${\rm [Fe/H]_{phot,v}}$ are satisfactory, comparable to many of the most recent examples found in the literature \citep[see, e.g.,][and references therein]{yang22,chandra22}. 

In the range ${\rm [Fe/H]_{spec}}\le -0.8$ there is a small asymmetry in the distribution of $\Delta {\rm [Fe/H]}$, with a sparse set of stars on the negative side, beyond the bulk of well-behaved stars. The amplitude of the discrepancy is large only for a  handful of them: 37 of the 1804 stars with  ${\rm [Fe/H]_{spec}}\le -0.8$ have $\Delta {\rm [Fe/H]}<-0.5$~dex, while those having $\Delta {\rm [Fe/H]}>+0.5$~dex in the same metallicity range are just six. For these stars the corrected ${\rm [Fe/H]_{phot,v}}$ overestimates ${\rm [Fe/H]_{spec}}$ by a significant amount. For the stars around 
${\rm [Fe/H]}\simeq -1.0$ this is possibly due to anomalously strong CN features in their spectra. On the other hand the asymmetry toward negative residuals for ${\rm [Fe/H]}\la -1.5$ is mainly due to stars close to the blue edge of the assumed validity range of the C07 calibration, $(v-y)_0\la 1.2$, where the sensitivity of $m_{1,0}$ to metallicity is maximal. However this comparison, as well as those presented in Appendix~\ref{appe_val}, suggest that this kind of problem affects only a few per cent of metal poor stars.

The typical precision of ${\rm [Fe/H]_{phot,b}}$ is significantly worse than ${\rm [Fe/H]_{phot,v}}$, at any metallicity. However it may be a useful tool in the metal-rich range beyond the applicability range of the C07 calibration, where it provides a higher median accuracy than ${\rm [Fe/H]_{phot,v}}$, as well as a weaker dependency on age. 

Given the results and discussion above, we define our final photometric metallicities as follows:

\begin{equation}
\label{eq:fe}
    {\rm [Fe/H]_{C} = [Fe/H]_{phot,v} +0.14}
\end{equation}
and
\begin{equation}
\label{eq:efe}
    {\rm \epsilon [Fe/H]_{C} = \epsilon[Fe/H]_{phot,v}}
\end{equation}

\noindent
for the C07 calibration;

\begin{equation}
\label{eq:fe_h}
    {\rm [Fe/H]_{H} = [Fe/H]_{phot,b} -0.36}
\end{equation}
and
\begin{equation}
\label{eq:efe_h}
    {\rm \epsilon [Fe/H]_{H} = \epsilon[Fe/H]_{phot,b}}
\end{equation}

\noindent
for the H00 calibration

In this way our photometric metallicities are calibrated on the APOGEE DR17 scale. These are the metallicity values, together with their uncertainties, that we provide in the publicly released dataset. The $(v-y)_0$ cuts adopted in Sect.~\ref{sample} include in the final sample 37020 stars (5\% of the entire dataset) having
$(b-y)_0$ colours slightly bluer that the validity limit of the H00 calibration, down to $(b-y)_0=0.43$, where the blue limit is
$(b-y)_0=0.50$. To facilitate the identification of  stars whose ${\rm [Fe/H]_H}$ are actually extrapolated, we provide also a flag ({\tt flagH}) that is valued 1 for stars bluer than $(b-y)_0=0.50$ and 0 otherwise.

The first rows of the dataset are reported in Table~\ref{tab:sample}, as a sample of its content. 
In the following analysis, focused on metal-poor populations, we will always use metallicity from the C07 calibration (${\rm [Fe/H]_C}$).

%###################################################################
\begin{figure*}[!ht]
\includegraphics[scale=0.23]{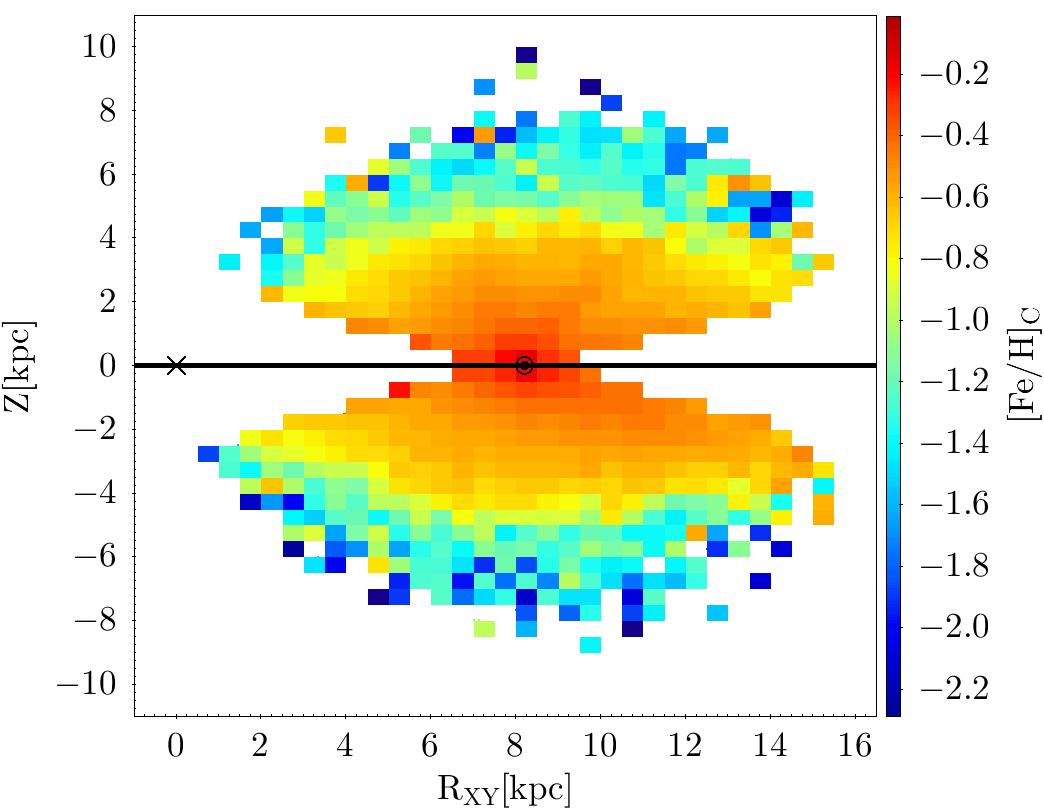}
\hspace{0.4cm}
\includegraphics[scale=0.23]{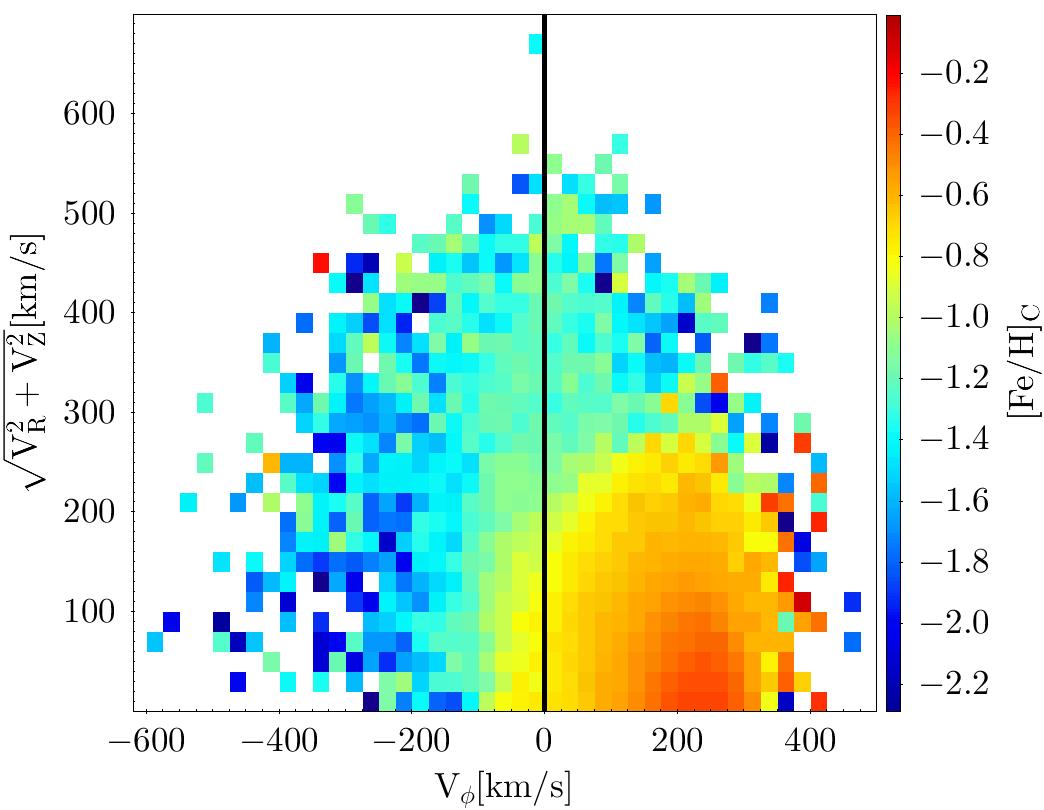}
\caption{Location in the Galaxy and kinematics of the stars in our sample. Left panel: bi-dimensional histogram in the $Z$ vs. $R_{XY}=\sqrt{X^2+Y^2}$ plane. 
The position of the Sun is indicated by the $\sun$ symbol, while the Galactic center is denoted by a  $\times$ symbol. 
Right panel: Toomre's diagram as a bi-dimensional histogram 
of the 685087 stars in our sample having a radial velocity measure from \Gaia DR3. In both panels the pixels (histogram bins) are colour-coded according to the median metallicity of stars in the bin.
\label{fig:toomre}
}
\end{figure*}
%###################################################################

\section{Chemo-kinematics of the sample}
\label{chemokin}

The main goal of large samples of stellar metallicities is to probe the chemo-dynamical properties of the stars in the various Galactic components. In this sense, an indirect test of the reliability and  power of our dataset is to verify if, and to what degree, the main correlations that are known to exist between position, kinematics and chemical composition are recovered. Figure~\ref{fig:toomre} shows that this is clearly the case\footnote{It is worth noting here that 685,087 out of 694233 stars in our sample (98.7\%) have valid line-of-sight velocity estimates from \Gaia DR3 \citep{dr3_rv}.}. The metallicity stratification with distance from the Galactic Plane (Z), with more metal-rich disc stars lying near the plane and progressively more-metal poor stars at larger Z, is apparent in the left panel of the figure. The Toomre diagram in the right panel of the figure shows the corresponding kinematic properties, with the most metal-rich stars showing high rotation, with a mode close to the assumed rotation velocity of the Sun 
\citep[$V_{\phi,\sun}=231$~km~s$^{-1}$;][]{mcmillan17}, 
and the metal-poor stars at higher $\sqrt{V_R^2+V_Z^2}$ velocities, also reaching highly retrograde orbits. 

Both plots are very similar to analogous diagrams obtained with spectroscopic samples \citep[see, e.g.,][and references therein]{helmi20,dr3_chimica}, demonstrating that our sample can indeed be fruitfully used to explore the properties of Galactic components and substructures. As an example, in Sect.~\ref{metasub} and Sect.\ref{mdfsub} we show how the metallicity distribution functions of known relics of past merging events can be investigated with unprecedentedly large samples.

\subsection{Metallicity of known sub-structures}
\label{metasub}

A natural application for our sample of stellar metallicities is related to Galactic archaeology, and in particular to the chemical characterisation of the dynamical sub-structures that have recently been discovered to populate the Milky Way Halo (see e.g. \citealt{helmi20} for a detailed review). These substructures are the phase-mixed debris of past merger events that the Milky Way experienced some Gyrs ago with now-disrupted dwarf galaxies, that still preserve coherency in the dynamical space described by the integrals of motion \citep{helmi00}.
To reconstruct the properties of these dwarf galaxies, a key ingredient comes from their MDFs. So far, this information has been gathered by pairing the dynamical information from \Gaia with the spectroscopic measurements provided by large surveys such as APOGEE \citep{apogee16}, GALAH \citep{galah_dr3}, LAMOST \citep{lamost}, RAVE \citep{rave}, and \Gaia DR3 \citep{dr3_chimica}. In this way, though, the availability of spectroscopic information has always played the limiting role, and the MDFs of the different substructures have typically been constructed using some hundreds of stars at best \citep{helmi18, koppelman19, naidu20, myeong22}, while the dynamical information is available for many thousands of stars.

\begin{figure}[ht!]
\center{
\includegraphics[width=\columnwidth]{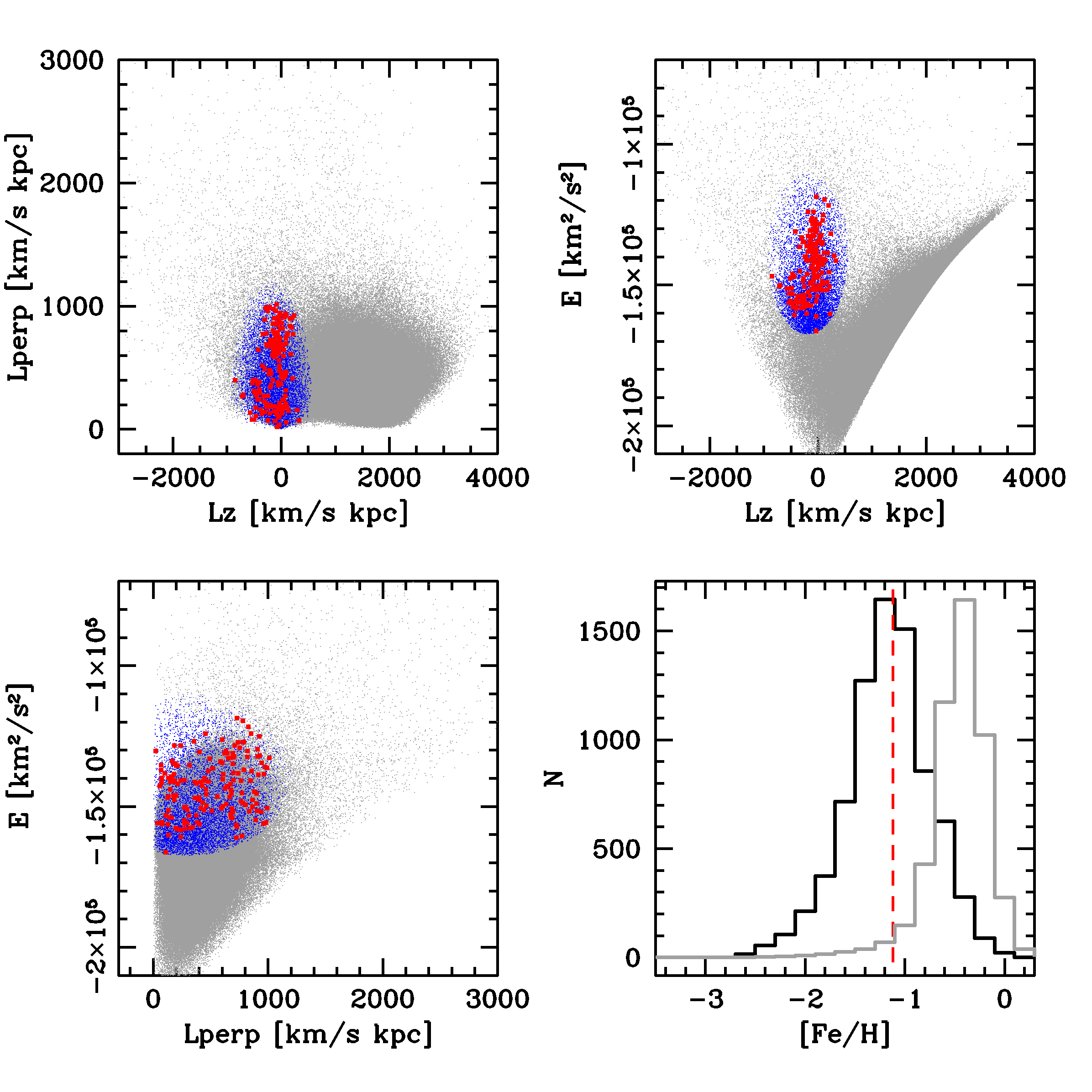}
}
\caption{Example of sub-structure selection. The upper panels and the lower-left panel show the location of the whole sample (grey dots) in the dynamical spaces described by L$_z$, L$_{perp}$ and E. The red thick dots mark stars in common with the list of GES stars from \citet{lovdal22, ruizlara22}. The blue symbols indicate our selection of likely GES members. The lower-right panel shows the MDF for this selection of GES stars (black histogram) and that for the entire sample, normalised to the peak of the former (grey histogram). The red-dashed line marks the mean [Fe/H] quoted by \citet{ruizlara22} for their own GES selection.  }
\label{fig:iom_ges}
\end{figure}

The method proposed in this paper allows us to overcome this limitation, obtaining metallicity information for the entirety of our \Gaia-based sample of stars. Therefore, with the aim of characterising the MDFs of the most prominent substructures populating the Milky Way Halo, we first computed the orbital parameters for each of the 694233 stars of our catalogue. To do so, we used the script described in detail and made publicly available by \citet{vasiliev21}, which in turn is based on the AGAMA library for Galactic dynamics \citep{vasiliev19}. Orbits were computed by assuming the static Milky Way potential by \citet{mcmillan17}. The result of the orbital integration is summarised in Fig.\ref{fig:iom_ges}, where the stars in our sample are plotted as grey dots in the L$_z$-L$_{perp}$, L$_z$-E, and L$_{perp}$-E planes (L$_z$ and L$_{perp}$ being the vertical and the perpendicular components of the angular momentum, respectively, and E being the orbital energy).

In order to select stellar members of the most significant merger events experienced by the Milky Way, we relied on the dynamical classification of a sample of halo stars within 2.5 kpc proposed by \citet{lovdal22} and further discussed in \citet{ruizlara22} (hereafter RL22). Since these authors used a different assumption on the shape of the Galactic potential compared to ours, to identify each RL22 substructure we first looked for stars in common between our and their catalogue, then we identified the location of each RL22 substructure in the dynamical spaces as computed by us, and finally we performed an elliptical selection designed to include all of the members of the RL22 substructures. An example of such a procedure is summarised in Fig.\ref{fig:iom_ges} for the Gaia-Enceladus-Sausage \citep[GES][]{helmi18, belokurov18} merger event. Stars shown as red symbols are those labelled as GES stars by RL22 and in common with our catalogue. Blue symbols indicate our final selection of GES members among the 694233 giants subject of this study. The bottom-right panel of Fig.\ref{fig:iom_ges} shows the MDF for the selected GES members (black histogram), that is significantly more metal-poor on average than the distribution for the entire sample, normalised to the peak of the former (grey histogram).

\begin{figure}[ht!]
\center{
\includegraphics[width=\columnwidth]{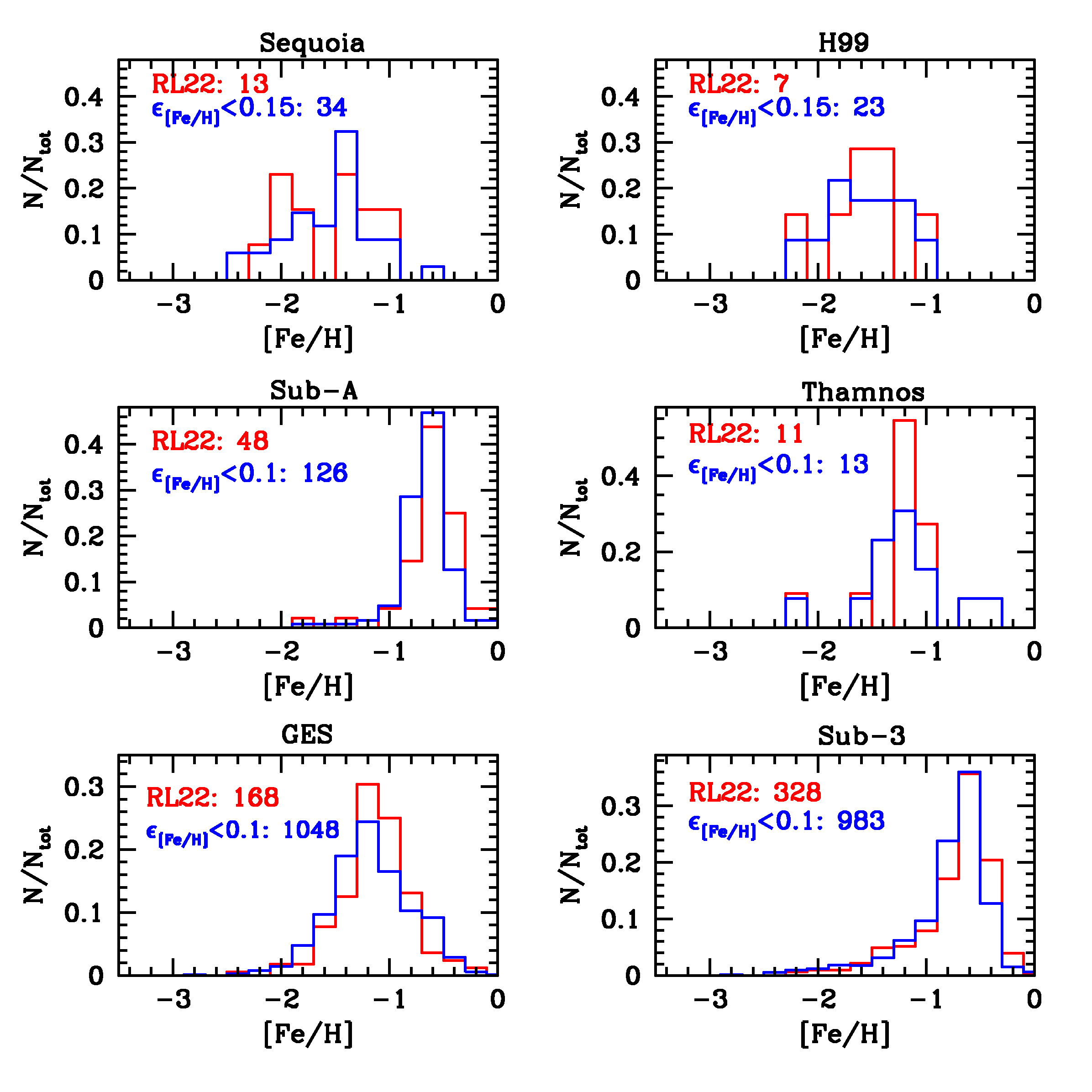}
}
\caption{Normalised MDFs for the six substructure under study, labelled at the top of each panel. The red histograms show the results obtained when only using stars in common with RL22, whereas the blue histograms show the distributions obtained with our own selection. The total number of stars used to build them is quoted in each panel, and listed in Table~\ref{tab:mdf}.}
\label{fig:mdfs}
\end{figure}

We followed the same procedure to select five other substructres. These are Sequoia \citep{myeong19}, the Helmi streams \citep[H99,][]{helmi99}, Thamnos \citep{koppelman19b}, RL22 substructure-A and substructure-3 (sub-A and sub-3 hereafter, respectively). The other substructures (or clusters) listed in RL22 have no stars in common with our sample, possibly because they limit their analysis to a sample of stars within a distance of 2.5 kpc.
Figure~\ref{fig:mdfs} shows the comparison between the MDF (normalised to the total number of involved stars) for each of the substructures as obtained when using the stars in common with RL22 (red histograms), and when using our sample of metallicities (blue histograms), after the application of a quality cut on $\epsilon{\rm [Fe/H]}$ (see the labels). The similarity between the results obtained with the two different approaches is remarkable, with the MDFs of the extended samples based on selections analogous to that illustrated in Fig.~\ref{fig:iom_ges} nicely reproducing those obtained from the original labelling by RL22, both in terms of peak location and metallicity range. 

Table \ref{tab:mdf} summarises the main properties of these MDFs in terms of 50th, 16th and 84th percentile. In good agreement with what has already been shown in the literature, sub-A and sub-3 are on average the most metal-rich. This is not surprising given that sub-A effectively corresponds to the hot thick disk (RL22), and that sub-3 \citep[also named L-RL3 in the most recent analysis by][]{dodd23} is dominated by a similar disky population and additionally shows an extended metal-poor tail of likely accreted stars that is also evident from our findings. On the other hand, Sequoia and H99 are the most metal-poor ones. 

Sub-A shows the tightest MDF, with  $\sigma_{[Fe/H]}^{sub-A}=0.31$~dex, GES, Sequoia and Thamnos have the broadest MDFs, all with $\sigma_{[Fe/H]}\simeq 0.8$. While the still low number of members might explain this large value for Sequoia and Thamnos, this feature fits the fact that the GES progenitor was the most massive, among these.

\begin{table}[!htbp]
%\centering
\caption{\label{tab:mdf} Characteristics of the high-quality MDFs of the six substructures, expressed in terms of 50th, 16th and 84th percentile.}
{
    \begin{tabular}{lccccr}
Substr. &  $P_{50}^{[Fe/H]}$  &  $P_{16}^{[Fe/H]}$  &   $P_{84}^{[Fe/H]}$ &  N$_{high-quality}$ & N$_{tot}$  \\  
\hline  
GES     & -1.19 & -1.52 &   -0.76 & 1048 &	7782  \\
Sub-3   & -0.70 & -1.09 &   -0.51 &  983 &	5998  \\
Sub-A   & -0.64 & -0.81 &   -0.50 &  126 &  896   \\
Thamnos & -1.22 & -1.46 &   -0.66 &  13  &  158   \\
Sequoia & -1.47 & -1.99 &   -1.16 &  34  &  64   \\
H99     & -1.52 & -1.87 &   -1.21 &  23  &  40   \\
\hline
    \end{tabular}
}
\tablefoot{Only stars satisfying the selection criteria displayed in Fig.~\ref{fig:mdfs} are considered in the computation of the reported percentiles. }
\end{table}

\subsection{Discussion}
\label{mdfsub}

\begin{figure*}[ht!]
\center{
\includegraphics[width=\textwidth]{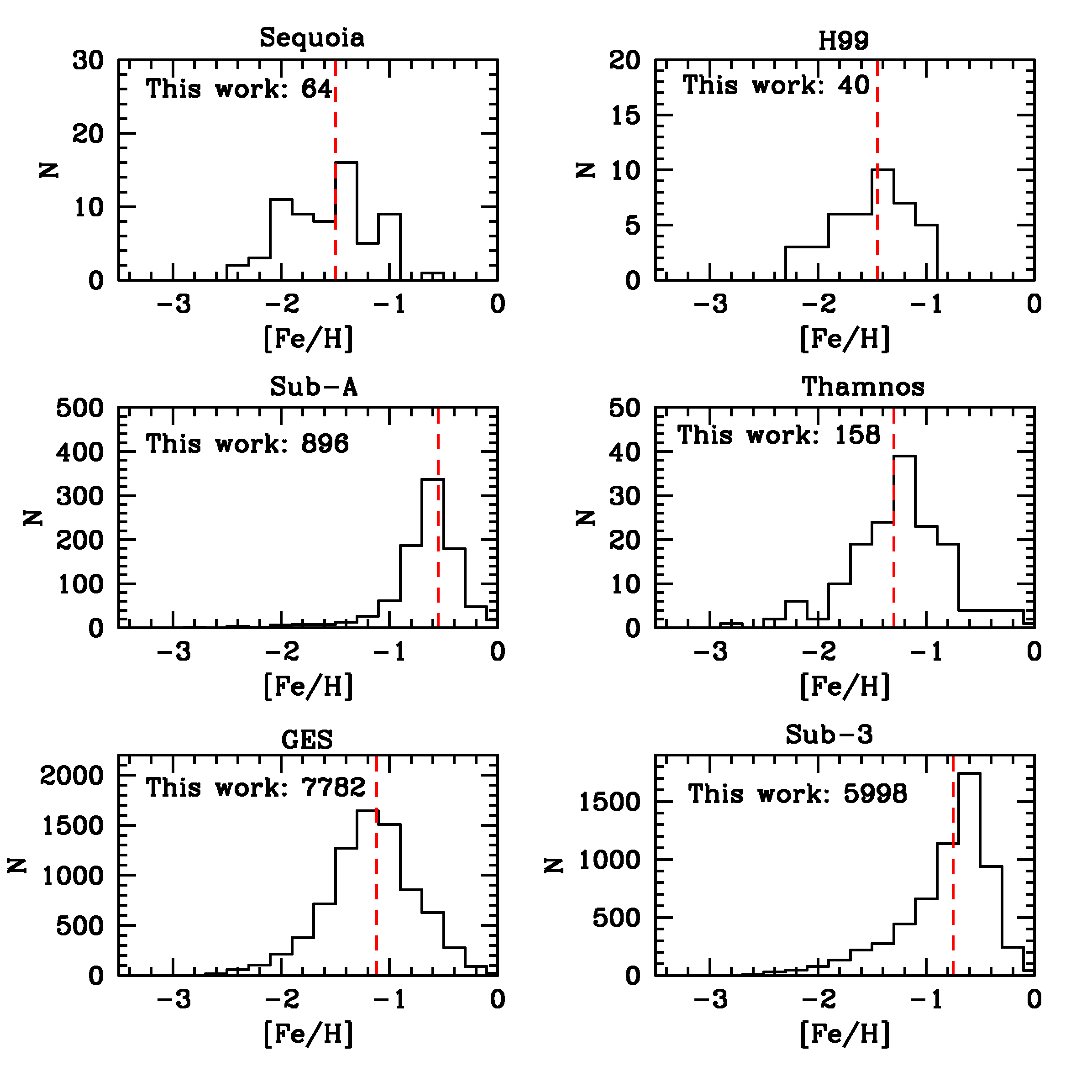}
}
\caption{MDFs for the six substructures. The red dashed line marks the metallicity of the best-fit theoretical model used by RL22 to fit the CMD of each substructure, and is shown as reference.}
\label{fig:mdfs_all}
\end{figure*}

Once we have demonstrated with a high-quality sub-sample that our selection produces MDFs that match well those obtained from selections found in the literature, we can investigate the MDFs of the six sub-structures when built with no quality selection whatsoever. The results are shown in Fig.\ref{fig:mdfs_all}. The overall shape of the MDF of each substructure is preserved, but the number of members in this case increases to reach several thousands of stars in the two most populous cases, that are GES and sub-3. 

By inspecting Fig.~\ref{fig:mdfs_all} it is very interesting to note that (a) all the MDF except the one of Sequoia display very clear modes, as well as shapes resembling those of existing Milky Way satellites \citep[see, eg.,][]{kirby11,hasse21}, (b) the mean and mode of the derived MDFs are in good agreement with analogous estimates obtained from (much smaller) spectroscopic samples \citep[e.g.,][]{koppelman19b, naidu20, horta22, ruizlara22}, (c) that the multi-modal MDF of Sequoia matches previous findings \citep[e.g.,][]{naidu20, monty20, horta22} that lead to the suggestion that the stars included in the Sequoia bin come from two or three independent structures that happen to lie close-by in the space of integral of motion (see RL22).

We note that all the considered sub-structures have relatively metal-rich mode (and median, see Tab.~\ref{tab:mdf}), e.g., with respect to the MDF of dwarf spheroidals like Draco, Ursa Minor, and Sextans \citep{kirby11}. GES and Thamnos have median metallicity between Leo~I (${\rm [Fe/H]_{med}=-1.42}$) and Fornax (${\rm [Fe/H]_{med}=-1.01}$), two dwarf spheroidals with total mass ranging from $1.2\times 10^7~M_{\sun}$ and $5.6\times 10^7~M_{\sun}$
\citep{mc12}.

The remarkably wide MDF of Sub-3 displays a strong peak at [Fe/H]$\simeq -0.7$, significantly more metal-rich than the mode of the GES MDF, with an extended metal-poor tail that reaches [Fe/H]$<-2.5$. As shown in RL22 and \citep{dodd23}, stars in this metal-poor component display chemical abundances of elements like Al, Mg and Mn that are typical of accreted stars, while the dominant metal-rich component shows chemistry consistent with that of an in-situ population. Our selection includes $516$ sub-3 stars having [Fe/H]$<-1.5$, which could be ideal targets for a high-resolution spectroscopic follow-up to better understand this possibly accreted component. This shows how powerful the metallicities derived with the method discussed in this paper are, and opens new opportunities in the investigation and characterisation of these merger events.

\begin{figure}[ht!]
\center{
\includegraphics[width=\columnwidth]{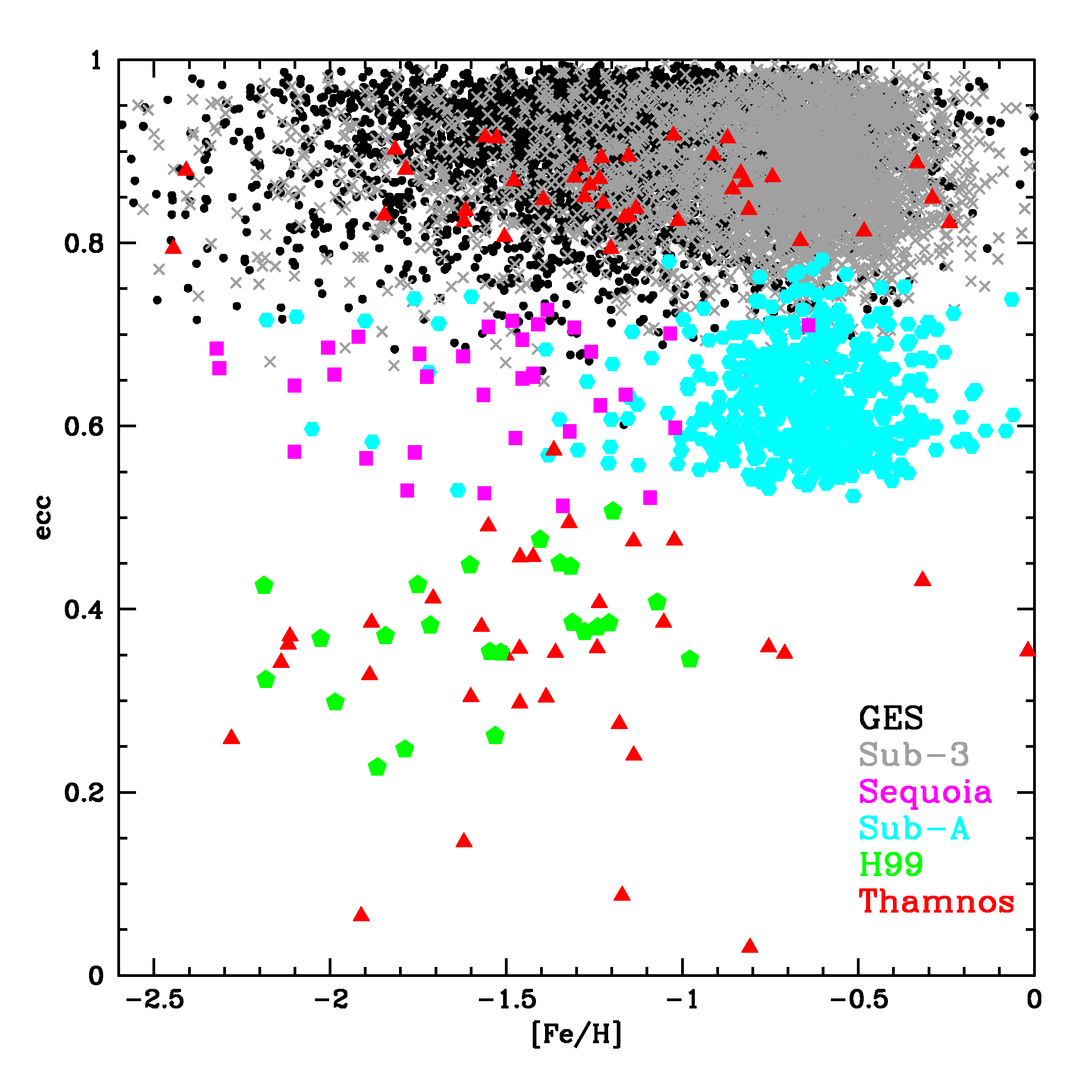}
}
\caption{Distribution of the six substructures in the chemo-dynamical plane described by [Fe/H] vs orbital eccentricity. }
\label{fig:ecc}
\end{figure}

\begin{figure}[ht!]
\center{
\includegraphics[width=\columnwidth]{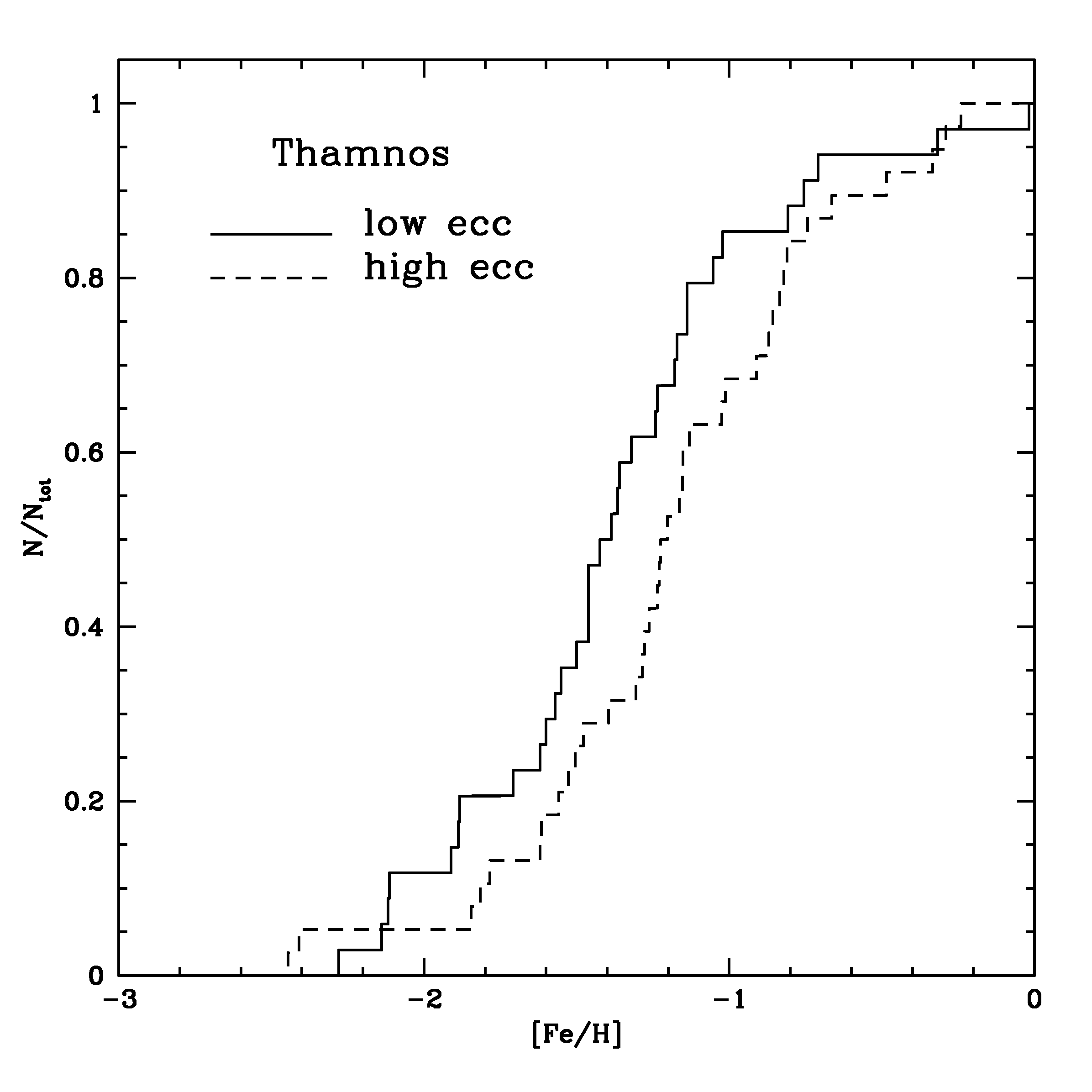}
}
\caption{Comparison between the cumulative metallicity distributions of the low-eccentricity (solid line) and the high-eccentricity (dashed line) populations of Thamnos. }
\label{fig:thamnosKS}
\end{figure}

One simple example in this sense is provided when combining the metallicity information with a dynamical quantity like the orbital eccentricity, expressed as $ecc=(apo-peri)/(apo+peri)$, where $apo$ and $peri$ are the orbital apocenter and pericenter, respectively. Figure~\ref{fig:ecc} shows how well this pair of parameters (plotted for stars with $\epsilon_{[Fe/H]}<0.15$) is able to distinguish the six substructures. Another interesting feature stemming out from this plane is that Thamnos stars (red triangles) are evidently split in two populations having significantly different eccentricity, one located at $ecc\simeq0.4$ and the other at $ecc\simeq0.85$. Very likely these two populations correspond to the two components that are already known to build Thamnos up, called Thamnos-1 and Thamnos-2 respectively \citep{koppelman19b}. This interpretation is supported by the fact that the population with lower eccentricity, which should correspond to Thamnos-1, is in fact more metal-poor on average ([Fe/H]$\simeq-1.4$) than the one at higher eccentricity ([Fe/H]$\simeq-1.15$), as already found in the discovery paper by \citet{koppelman19b}. However, in this context, comparing MDFs should provide more insight than comparing just means. This is done in Fig.~\ref{fig:thamnosKS} where it can be appreciated that Thamnos-2 MDF is indeed more metal-rich than that of Thamnos-1. A Kolmogorov-Smirnov test states that the probability for the two distributions to have been drawn from the same population is only 6\%.

\section{Summary and conclusions}
\label{conclu}

We used Str\"omgren $vby$ synthetic photometry from \Gaia DR3 XP spectra \citep{dpacp93} and the calibrations by C07 and H00, based on the $m_{1,0}$ index, to derive photometric metallicities for 694233 selected Galactic old giant stars. These calibrations are generally believed to provide age-independent metallicities \citep[see, e.g.,][]{dirsch00,narloch22}. A simple analysis performed with theoretical isochrones confirms that indeed the age dependency of $m_{1,0}$ is negligible, at least for ${\rm [Fe/H]}\la -0.5$.

A subset of 44782 stars in common with the APOGEE DR17 sample \citep{apogee_dr17} were used to re-calibrate the  photometric metallicities on this spectroscopic scale, by means of shifts that minimises the median difference $\Delta {\rm [Fe/H]}$ in the metallicity regime where no significant trend between $\Delta {\rm [Fe/H]}$ and ${\rm [Fe/H]_{spec}}$ is present.
This occurs in the range $-2.0\la {\rm [Fe/H]}\la -0.4$ for the metallicities derived from the C07 calibration (${\rm [Fe/H]_{C}}$), broadly corresponding with its validity range, and  ${\rm [Fe/H]}\ga -0.8$ for those from the H00 calibration (${\rm [Fe/H]_{H}}$). 

${\rm [Fe/H]_{C}}$ reproduces the spectroscopic APOGEE DR17 metallicity with a median accuracy $\le 0.10$~dex in the range $-2.0\le {\rm [Fe/H]_{spec}}\le -0.4$. At higher metallicity, beyond the applicability limit of the relation, a monotonic trend with ${\rm [Fe/H]_{spec}}$ arises, reaching a median amplitude of $0.50$~dex at  ${\rm [Fe/H]_{spec}}=+0.4$, with ${\rm [Fe/H]_{C}}$ systematically under-estimating ${\rm [Fe/H]_{spec}}$. This trend may be related to the onset of a significant age dependency arising in the high metallicity regime (see Sect.~\ref{cal} and Appendix~\ref{appe_trend}). The precision, as measured by $\sigma=0.5(P_{84}-P_{16})$, is $<0.25$~dex over the entire metallicity range considered, $<0.20$~dex for ${\rm [Fe/H]_{spec}}\ge -1.60$, and $\le 0.16$~dex for ${\rm [Fe/H]_{spec}}\ge -1.20$. 

 ${\rm [Fe/H]_{H}}$ reproduces the spectroscopic APOGEE DR17 metallicity with a median accuracy $\le 0.05$~dex in the range $-0.8\le {\rm [Fe/H]_{spec}}\le +0.4$, while, at lower metallicities, it displays trends with ${\rm [Fe/H]_{spec}}$ reaching amplitudes $> 0.20$~dex. The precision is nearly everywhere $\ga 0.2$~dex, reaching values $\ge 0.3$~dex toward the metal-poor limit. Theoretical and empirical arguments suggest that the age-dependency of ${\rm [Fe/H]_{H}}$ is weaker than that affecting ${\rm [Fe/H]_{C}}$, in the ${\rm [Fe/H]}>-0.5$ regime. However, since we are mainly interested to the metal-poor regime and given the higher precision, we take ${\rm [Fe/H]_{C}}$ as our preferred metallicity indicator, tailoring the sample selection on the properties of the C07 calibration.

We provide a publicly available catalogue with the re-calibrated photometric metallicities for all the stars in our sample together with the associated uncertainties. Our final metallicity values have been validated by comparison with large samples of spectroscopic metallicities from various surveys, with chemical abundances derived from spectra of different resolution (GALAH, \Gaia DR3 GSP-spec, LAMOST, and Gaia-ESO, see Appendix~\ref{appe_val}). The overall performances are remarkably similar to those described above for the comparison with APOGEE DR17, implying that our photometric metallicity scales are robust and have general validity. 
We have demonstrated that while individual values may be significantly inaccurate, the overall dataset traces the spectroscopic metallicities with sufficient accuracy and precision to allow useful scientific applications. 

Once the intrinsic uncertainties and applicability range of the adopted calibrating relation are taken into account, as well as those associated to the spectroscopic metallicity scale taken as reference, there are three main factors that hamper the accuracy and the precision of the metallicities derived here: uncertainties on interstellar extinction, either statistic or systematic, photometric precision, and spectral features mimicking the effect of metallicity, like, e.g. the well known effect of CN bands (H00). Improvements in the first two factors can be foreseen for the next \Gaia data releases.

In Sect.~\ref{chemokin} we showed how well our photometric metallicity is able to trace the chemo-kinematic trends of the stars included in our dataset. This was conducive to derive the most richly populated MDFs ever presented in the literature of several known substructures in the surroundings of the Sun (GES, Thamnos, Sequoia, Sub-3, Sub-A, H99), in some case providing precious insights on the actual nature of the considered substructures and/or their progenitors.

This work provides a simple and tested way to get photometric metallicity of old giants from standardised Str\"omgren XP synthetic photometry. The method adopted here can be extended to larger samples of Galactic stars or applied to stellar systems with known distance and extinction. Finally, it provides an additional proof that very useful astrophysical information is encoded in \Gaia XP spectra \citep{dr3_spectra_fda,dr3_ec_spectra} and that it can be efficiently extracted using synthetic photometry \citep{dpacp93}.  

\begin{acknowledgements}

MB, PM, AB and DM acknowledge the support to activities related to the ESA/\Gaia mission by the Italian Space Agency (ASI) through contract 2018-24-HH.0 and its addendum 2018-24-HH.1-2022 to the National Institute for Astrophysics (INAF). MB, AB and DM acknowledge the support to this study by the PRIN INAF 2019 grant ObFu 1.05.01.85.14 ({\em Building up the halo: chemo-dynamical tagging in the age of large surveys}, PI. S. Lucatello). MB is grateful to R. Pascale for his help in the production of figures with Python.

This work has made use of data from the European Space Agency (ESA) mission \Gaia (https://www.cosmos.esa.int/Gaia), processed by the \Gaia Data Processing and Analysis Consortium (DPAC, https://www.cosmos.esa.int/web/Gaia/dpac/consortium). Funding for the DPAC has been provided by national institutions, in particular the institutions participating in the \Gaia Multilateral Agreement.

In this analysis we made use of TOPCAT (http://www.starlink.ac.uk/topcat/, \citealt{Taylor2005}).

This work made use of SDSS-IV data. Funding for the Sloan Digital Sky Survey IV has been provided by the 
Alfred P. Sloan Foundation, the U.S. Department of Energy Office of Science, and the Participating 
Institutions. 
SDSS-IV acknowledges support and resources from the Center for High Performance Computing  at the 
University of Utah. The SDSS website is www.sdss.org.
SDSS-IV is managed by the Astrophysical Research Consortium for the Participating Institutions of the SDSS Collaboration including the Brazilian Participation Group, the Carnegie Institution for Science, Carnegie Mellon University, Center for Astrophysics | Harvard \& Smithsonian, the Chilean Participation Group, the French Participation Group, Instituto de Astrof\'isica de Canarias, The Johns Hopkins University, Kavli Institute for the Physics and Mathematics of the Universe (IPMU) / University of Tokyo, the Korean Participation Group, Lawrence Berkeley National Laboratory, Leibniz Institut f\"ur Astrophysik Potsdam (AIP),  Max-Planck-Institut f\"ur Astronomie (MPIA Heidelberg), Max-Planck-Institut f\"ur Astrophysik (MPA Garching), 
Max-Planck-Institut f\"ur Extraterrestrische Physik (MPE), National Astronomical Observatories of 
China, New Mexico State University, New York University, University of Notre Dame, Observat\'ario 
Nacional / MCTI, The Ohio State University, Pennsylvania State University, Shanghai 
Astronomical Observatory, United Kingdom Participation Group, Universidad Nacional Aut\'onoma de M\'exico, University of Arizona, University of Colorado Boulder, University of Oxford, University of Portsmouth, University of Utah, University of Virginia, University of Washington, University of Wisconsin, Vanderbilt University, and Yale University.

This work made use of the Third Data Release of the GALAH Survey (Buder et al. 2021). The GALAH Survey is based on data acquired through the Australian Astronomical Observatory, under programs: A/2013B/13 (The GALAH pilot survey); A/2014A/25, A/2015A/19, A2017A/18 (The GALAH survey phase 1); A2018A/18 (Open clusters with HERMES); A2019A/1 (Hierarchical star formation in Ori OB1); A2019A/15 (The GALAH survey phase 2); A/2015B/19, A/2016A/22, A/2016B/10, A/2017B/16, A/2018B/15 (The HERMES-TESS program); and A/2015A/3, A/2015B/1, A/2015B/19, A/2016A/22, A/2016B/12, A/2017A/14 (The HERMES K2-follow-up program). We acknowledge the traditional owners of the land on which the AAT stands, the Gamilaraay people, and pay our respects to elders past and present. This paper includes data that has been provided by AAO Data Central (datacentral.org.au).

This work made use of LAMOST data.
Guoshoujing Telescope (the Large Sky Area Multi-Object Fiber Spectroscopic Telescope LAMOST) is a National Major Scientific Project built by the Chinese Academy of Sciences. Funding for the project has been provided by the National Development and Reform Commission. LAMOST is operated and managed by the National Astronomical Observatories, Chinese Academy of Sciences.

This work made use of Gaia-ESO Public Spectroscopic Survey data products from observations made with the ESO Very Large Telescope at the La Silla Paranal Observatory under programme ID 188.B-3002. These data products have been processed by the Cambridge Astronomy Survey Unit (CASU) at the Institute of Astronomy, University of Cambridge, and the FLAMES/UVES reduction team at INAF/Osservatorio Astrofisico di Arcetri. The Gaia-ESO Survey Data Archive is prepared and hosted by the Wide Field Astronomy Unit, Institute for Astronomy, University of Edinburgh, which is funded by the UK Science and Technology Facilities Council.

\end{acknowledgements}

%\input{sections/outputs.tex}
%\input{sections/recommendations.tex}
%\input{sections/conclusions.tex}

%--------------------------------------------------------------------

\bibliographystyle{aa} % style aa.bst
\bibliography{refs} % your references refs.bib

\begin{appendix}

\section{Validation with GALAH, GSP-Spec, LAMOST and Gaia-ESO spectroscopic metallicities}
\label{appe_val}

Here we show the same kind of analysis performed in Sect.~\ref{apogee} for four additional VSs, composed of stars with well-measured ${\rm [Fe/H]_{spec}}$ from the GALAH \citep{galah}, LAMOST \citep{lamost},  Gaia-ESO \citep{ges_field,ges_open} surveys, and from the GSP-Spec set of chemical abundances derived from RVS spectra for \Gaia DR3 \citep{gspspec}. 

In Sect.~\ref{apogee} we used the APOGEE VS to calibrate our photometric metallicities, adjusting the zero points of our [Fe/H] scales to the APOGEE scale in the most appropriate metallicity ranges. 
Here, on the other hand, we compare our final calibrated metallicities, as defined in Eq.~\ref{eq:fe} and Eq.~\ref{eq:fe_h}, with large and reliable external spectroscopic samples, for validation purposes. It is important to recall that part of the trends or zero point differences in the median ${\rm \Delta [Fe/H]}$  as a function of ${\rm [Fe/H]_{spec}}$ that arise in the comparisons presented below may be due to non homogeneity between the abundance scale of APOGEE and that of the other surveys. The main result of the following analysis is that our photometric metallicities, calibrated on APOGEE data, nicely reproduce spectroscopic metallicity for the considered VS, with typical accuracy and precision similar to those described in Sect.~\ref{apogee}.

\subsection{Validation with GALAH data}

The GALactic Archaeology with HERMES (GALAH) survey \citep{galah15,galah} collects high-resolution ($R\simeq 28000$) optical spectra of Milky Way stars, deriving chemical abundances of many elements. From the third data release of the survey \citep{galah_dr3}, we selected for our VS only stars with good quality flags and small uncertainty in [Fe/H] and [$\alpha/$Fe]. The requirement on [$\alpha/$Fe] was motivated to allow the possibility of looking for trends of photometric metallicity with this abundance parameter. The issue is briefly discussed in Appendix~\ref{appe_trend}.  We extracted from {\tt GALAH\_DR3\_main\_allstar\_v2.fits} stars having {\tt flag\_sp=0}, {\tt flag\_fe\_h=0}, {\tt e\_fe\_h$<0.1$}, and {\tt e\_alpha\_fe$<0.2$} \footnote{See \url{https://www.galah-survey.org/dr3/overview/}}, finding 26286 of them in common with our sample.

%%%%%%%%%%%%%%%%%%%
\begin{figure}[ht!]
\center{
\includegraphics[width=\columnwidth]{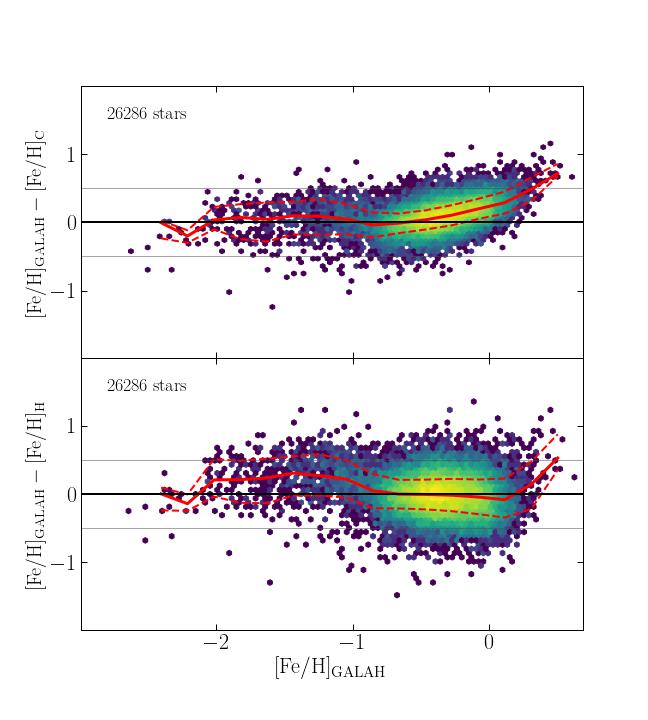}
}
\caption{${\rm \Delta [Fe/H]}$ for ${\rm [Fe/H]_C}$ (upper panel) and ${\rm [Fe/H]_H}$ (lower panel) as a function of ${\rm [Fe/H]_{spec}}$ for the GALAH VS. The arrangement and the meaning of the symbols is the same as Fig.~\ref{fig:apo_fe}.}
\label{fig:galah_val}
\end{figure} 
%%%%%%%%%%%%%%%%%%%%%%%%%%%%%%%%%%%%%%%%%

The distributions of ${\rm \Delta [Fe/H]_C}$ and ${\rm \Delta [Fe/H]_H}$ as a function of ${\rm [Fe/H]_{spec}}$ are shown in Fig.~\ref{fig:galah_val}. Median accuracy, precision and the overall behaviour of the distributions are very similar to those observed for the APOGEE VS.

\subsection{Validation with \Gaia DR3 GSP-Spec data}

With the third \Gaia data release, for the first time, chemical abundances were derived from the medium resolution spectra ($R\simeq 11500$) in a narrow range around the Calcium triplet ($845-872$~nm) from the RVS  spectrograph, with the GSP-Spec module \citep[][and references therein]{gspspec}. For our VS we selected best-measured stars by keeping only those with the 13 first bits
of {\tt astrophysical\_parameters.flags\_gspspec} equal to zero\footnote{{\tt flags\_gspspec LIKE ‘0000000000000\%’}}. This selection leaves only less than a hundred of stars with ${\rm [Fe/H]_{spec}}\la -1.0$\footnote{See \citet{gspspec} for a selection criterion retaining a significantly more conspicuous sample of metal-poor stars.}, but provides 224404 optimally measured stars in common with our sample. GSP-spec abundances were corrected for the gravity-dependent bias using Eq.~2 and Eq.~3 of \citet{gspspec}.

%%%%%%%%%%%%%%%%%%%
\begin{figure}[ht!]
\center{
\includegraphics[width=\columnwidth]{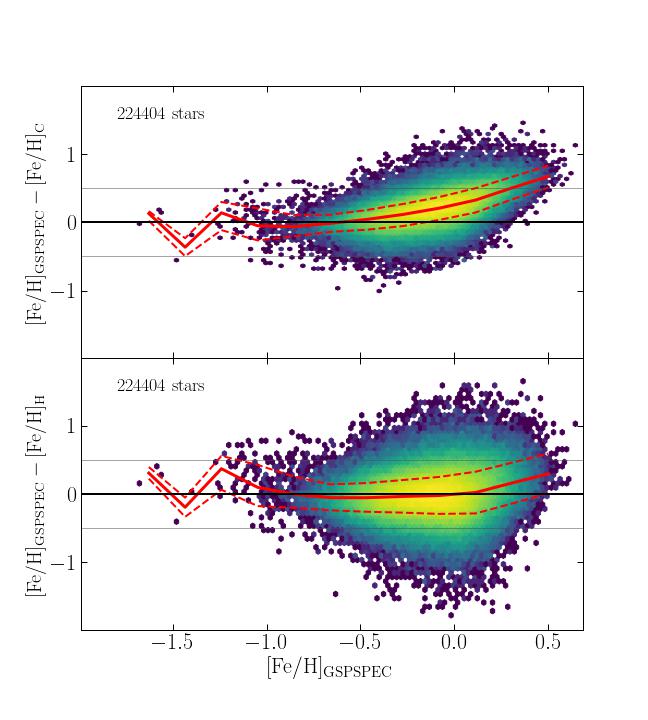}
}
\caption{ ${\rm \Delta [Fe/H]}$ for ${\rm [Fe/H]_C}$ (upper panel) and ${\rm [Fe/H]_H}$ (lower panel) as a function of ${\rm [Fe/H]_{spec}}$ for the GSP-Spec VS. The arrangement and the meaning of the symbols is the same as Fig.~\ref{fig:apo_fe}.}
\label{fig:gspspec_val}
\end{figure} 
%%%%%%%%%%%%%%%%%%%%%%%%%%%%%%%%%%%%%%%%%

Fig.~\ref{fig:gspspec_val} shows that our ${\rm [Fe/H]_C}$ values reproduce those by GSP-Spec with accuracy and precision similar to those observed for the APOGEE and GALAH VSs, in the range well sampled by the GSP-Spec VS. The behaviour of ${\rm [Fe/H]_H}$ is also broadly similar to that seen in previously considered VSs, however the precision is slightly worse everywhere and a trend with ${\rm [Fe/H]_{spec}}$ emerges for ${\rm [Fe/H]_{spec}}\ga 0.0$.

\subsection{Validation with LAMOST data}

To have a comparison with a VS with metallicities from low resolution spectra we selected stars from the 8th data release \citep[DR8;][]{lamost_dr8}
of the Large Sky Area Multi-object Fiber Spectroscopic Telescope (LAMOST) survey \citep[][]{lamost_tot}. The LAMOST catalogue collects abundances derived from spectra at $R\simeq 1800$ over the wavelength range 369-910~nm.

%%%%%%%%%%%%%%%%%%%
\begin{figure}[ht!]
\center{
\includegraphics[width=\columnwidth]{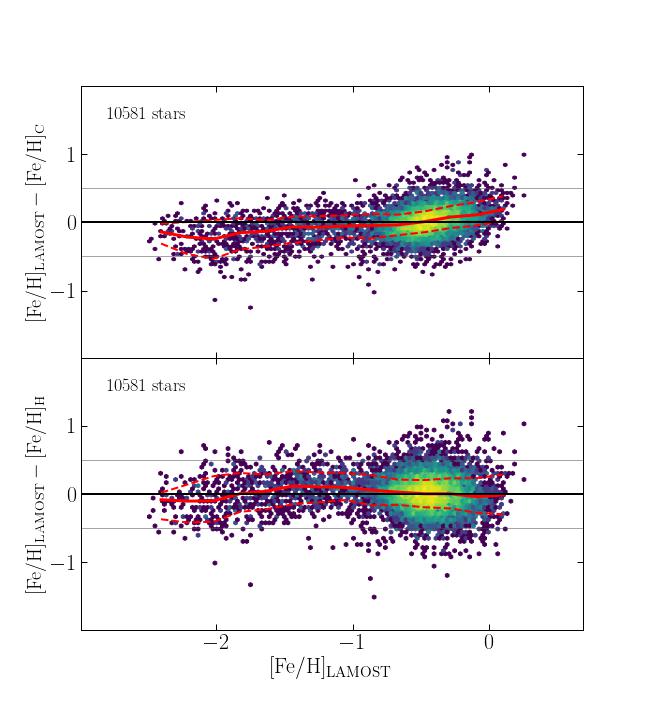}
}
\caption{${\rm \Delta [Fe/H]}$ for ${\rm [Fe/H]_C}$ (upper panel) and ${\rm [Fe/H]_H}$ (lower panel) as a function of ${\rm [Fe/H]_{spec}}$ for the LAMOST VS. The arrangement and the meaning of the symbols is the same as Fig.~\ref{fig:apo_fe}.}
\label{fig:lamost_val}
\end{figure} 
%%%%%%%%%%%%%%%%%%%%%%%%%%%%%%%%%%%%%%%%%

We extracted the best-measured sources for the LAMOST VS data from the DR8 table {\tt dr8\_v2.0\_LRS\_stellar.fits}, requiring that {\tt snrg>100}, {\tt feh\_err$\le$0.1}, {\tt alpha\_m\_err$\le$0.1}, and {\tt -1.0<alpha\_m<2.0}\footnote{See \url{http://www.lamost.org/dr8/}}. We found 10581 stars with this properties in common with our sample. For this VS, the distributions of ${\rm \Delta [Fe/H]}$ as a function of ${\rm [Fe/H]_{spec}}$ are displayed in Fig.~\ref{fig:lamost_val}. Concerning  the comparison with ${\rm [Fe/H]_C}$, as expected, $\sigma_{[Fe/H]}$ is slightly larger than what measured in the comparison with metallicity from high-resolution high-SNR spectra, owing to the larger uncertainties associated to LAMOST ${\rm [Fe/H]_{spec}}$. On the other hand the median accuracy is better than 0.15~dex over most of the considered metallicity range. Also in this case, the precision of ${\rm [Fe/H]_H}$ is worse than ${\rm [Fe/H]_C}$, but the median precision is $\le 0.12$ over the entire metallicity range sampled by the VS.

\subsection{Validation with Gaia-ESO data}

Gaia-ESO \citep{ges_field,ges_open} collected high resolution spectra of about 115000 stars in the MW field and in stellar clusters, using FLAMES at the ESO VLT. The subset of stars observed with UVES at resolution $R\simeq45000$ using the U580 setup (480-680 nm) provides the highest precision stellar parameters and abundances, so we concentrated on it. 
We used the final data release (available from the ESO catalogue archive\footnote{\url{https://www.eso.org/qi/catalogQuery/index/393}}, cross-matched it with \Gaia DR3, and kept only stars with U580 spectra. Then, using {\sc topcat} \citep{topcat}, we selected red giant stars based on the Kiel diagram (T$_{ eff}$, $\log g$) and applied a final cut on errors, keeping only stars
with errors in T$_{ eff}$, $\log g$, and [Fe/H]
less than 100~K, 0.25~dex, and 0.1~dex, respectively. This led to a sample of 1759 stars, 368 of which are in common with our sample. The small numbers of stars involved makes the distribution quite noisy, especially for [Fe/H]$\le -1.0$. Still, in all the bins with more than 10 stars the median ${\rm \Delta [Fe/H]_C}$ as well as $\sigma$ are $\la 0.25$~dex. As usual ${\rm \Delta [Fe/H]_H}$ displays lower precision at any metallicity and larger median deviations from zero in the metal-poor regime, while the trend with ${\rm [Fe/H]_{spec}}$ is virtually null for ${\rm [Fe/H]_{spec}}\ga -0.5$.

%%%%%%%%%%%%%%%%%%%
\begin{figure}[ht!]
\center{
\includegraphics[width=\columnwidth]{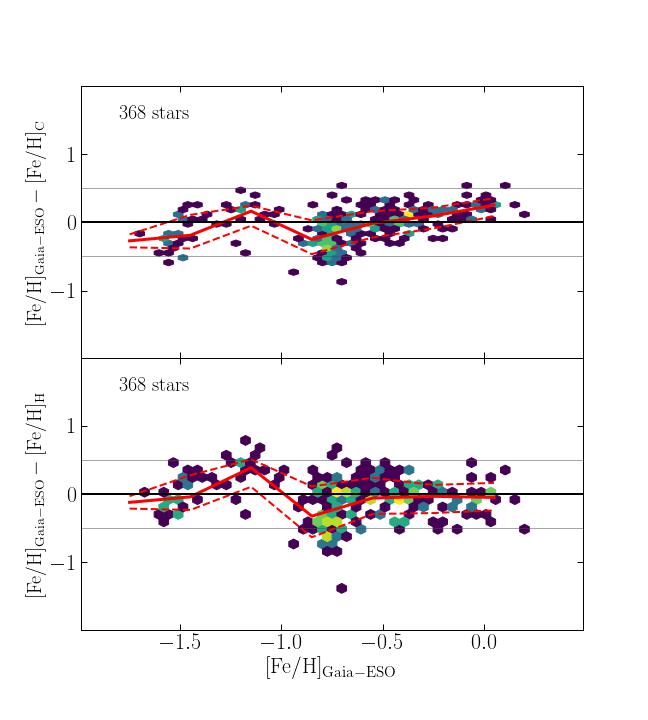}
}
\caption{ ${\rm \Delta [Fe/H]}$ for ${\rm [Fe/H]_C}$ (upper panel) and ${\rm [Fe/H]_H}$ (lower panel) as a function of ${\rm [Fe/H]_{spec}}$ for the Gaia-ESO VS. The arrangement and the meaning of the symbols is the same as Fig.~\ref{fig:apo_fe} but the median and percentile lines are computed in 0.3~dex bin, due to the small dimension of the sample.}
\label{fig:ges_val}
\end{figure} 
%%%%%%%%%%%%%%%%%%%%%%%%%%%%%%%%%%%%%%%%%

\section{Performances of C07 semiempirical calibrations}
\label{app_semi}

In addition to the empirical calibration we used here (Eq.~\ref{cal}), C07 provide also theoretical and semiempirical calibrations to obtain metallicity from Str\"omgren indices. Here we briefly explore the performance of C07 semiempirical calibrations against the APOGEE VS, since they have a wider range of applicability than those adopted here, extending to $-2.6 \le {\rm [Fe/H]} \le -0.6$, instead of $-2.2 \le {\rm [Fe/H]} \le -0.7$. Among C07 semiempirical calibrations, there is also one where the reddening-independent $[m]=m+0.3(b-y)$ index is used as metallicity indicator instead of $m_{1,0}$. Note, however, that metallicities derived in this way are not fully reddening-independent, as the calibrating relation depends on $[m]$ and on $(v-y)_0$. 

The semiempirical calibration as a function of $m_{1,0}$ and $(v-y)_0$ is:

\begin{equation}
    \label{eq:cal_semi}
{\rm [Fe/H]_{phot,s}}=\frac{m_{1,0}-0.521(v-y)_0+0.309}{0.159(v-y)_0-0.090}
\end{equation}

while the one as a function of $[m]$ and $(v-y)_0$ is:

\begin{equation}
    \label{eq:cal_mm}
{\rm [Fe/H]_{phot,[m]}}=\frac{[m]-0.585(v-y)_0+0.251}{0.131(v-y)_0-0.070}
\end{equation}

%%%%%%%%%%%%%%%%%%%
\begin{figure}[ht!]
\center{
\includegraphics[width=\columnwidth]{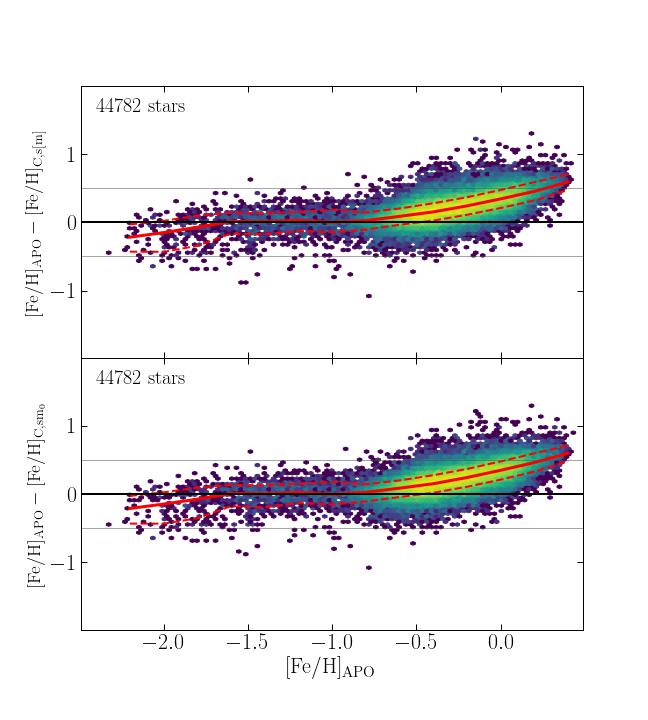}
}
\caption{${\rm \Delta [Fe/H]}$ for ${\rm [Fe/H]_{C,s[m]}}$ (upper panel) and ${\rm [Fe/H]_{C,sm_0}}$ (lower panel) as a function of ${\rm [Fe/H]_{spec}}$ for the APOGEE VS. The arrangement and the meaning of the symbols is the same as Fig.~\ref{fig:apo_fe}.}
\label{fig:apo_semi}
\end{figure} 
%%%%%%%%%%%%%%%%%%%%%%%%%%%%%%%%%%%%%%%%%

In Fig.~\ref{fig:apo_semi} we show the comparison of these photometric metallicities with ${\rm [Fe/H]_{APO}}$ after the following zero-point adjustments are performed, analogous to Eq.~\ref{eq:fe} and Eq.~\ref{eq:fe_h}:

\begin{equation}
\label{eq:fe_sm}
   {\rm [Fe/H]_{C,sm_0} = [Fe/H]_{phot,s} +0.10}  
\end{equation}

\noindent
and

\begin{equation}
\label{eq:fe_cm}
   {\rm [Fe/H]_{C,s[m]} = [Fe/H]_{phot,[m]} +0.10}  
\end{equation}

\noindent
to minimise the median $\Delta {\rm [Fe/H]}$ in the metal-poor regime. It is interesting to note that C07 finds similar shifts when comparing photometric metallicities from the same calibrating relations to spectroscopic metallicities for a sample of 85 field giants they use for validation. 

The behaviour of both photometric metallicities is qualitatively similar to ${\rm [Fe/H]_{C}}$. The precision is very similar at all metallicities, but the median accuracy is slightly worse both in the metal-poor and metal-rich regimes.
For example, at ${\rm [Fe/H]}=-2.0$, ${\rm |\Delta [Fe/H]|}$ is 0.08~dex for ${\rm [Fe/H]_{C}}$ and 0.15~dex for ${\rm [Fe/H]_{C,s[m]}}$; at ${\rm [Fe/H]}=-0.4$, ${\rm |\Delta [Fe/H]|}$ is 0.07~dex for ${\rm [Fe/H]_{C}}$ and 0.15~dex for ${\rm [Fe/H]_{C,s[m]}}$\footnote{The median accuracy of ${\rm [Fe/H]_{C,s[m]}}$ and 
${\rm [Fe/H]_{C,sm_0}}$ are virtually indistinguishable.}. 

Hence, ${\rm [Fe/H]_{C}}$ is the C07-based metallicity that reproduces most accurately ${\rm [Fe/H]_{APO}}$ for stars of the APOGEE VS, among those considered here. For this reason we decided to include only ${\rm [Fe/H]_{C}}$ in our final catalogue (Tab.~\ref{tab:sample}), together with ${\rm [Fe/H]_{H}}$.
However, Eq.~\ref{eq:fe_sm}, Eq.~\ref{eq:fe_cm} makes possible, for any interested reader, to derive both ${\rm [Fe/H]_{C,s[m]}}$ and 
${\rm [Fe/H]_{C,sm_0}}$ in a scale fully homogeneous to ${\rm [Fe/H]_{C}}$, based on APOGEE DR17, with 
Fig.~\ref{fig:apo_semi} providing the validation and illustrating the performances as a function of metallicity.

\section{Experiments with star clusters}
\label{app_gctest}

Star clusters are essentially composed of stars with the same age and metallicity. For this reason they may provide a useful playground to test the performances of our photometric metallicities in real and somehow challenging cases. The latter is especially true for globular clusters (GCs) as they are relatively distant and crowding may affect XP synthetic photometry of their stars in spite of the strict selection criteria adopted in our catalogue, especially at faint magnitudes \citep[see][]{dpacp93}. 

GC stars are rare in our sample, due to the adopted selections. Among the clusters with |b|$>20.0\degr$ and $D<10.0$~kpc, we found only six having more than five member stars included in our catalogue (see Table~\ref{tab:gctest}). We considered as members only stars having probability membership P$\ge 0.99$, as assigned in \citet{vasiliev21}. We adopted GC spectroscopic metallicities from the 2010 edition of the \citep{harris1996} catalogue, that are in the \citet{carretta09} metallicity scale. Since all the considered GCs have ${\rm [Fe/H]_{spec}}< -0.70$ we use only photometric metallicities from the C07 calibration.

%%%%%%%%%%%%%%%%%%%
\begin{figure}[ht!]
\center{
\includegraphics[width=\columnwidth]{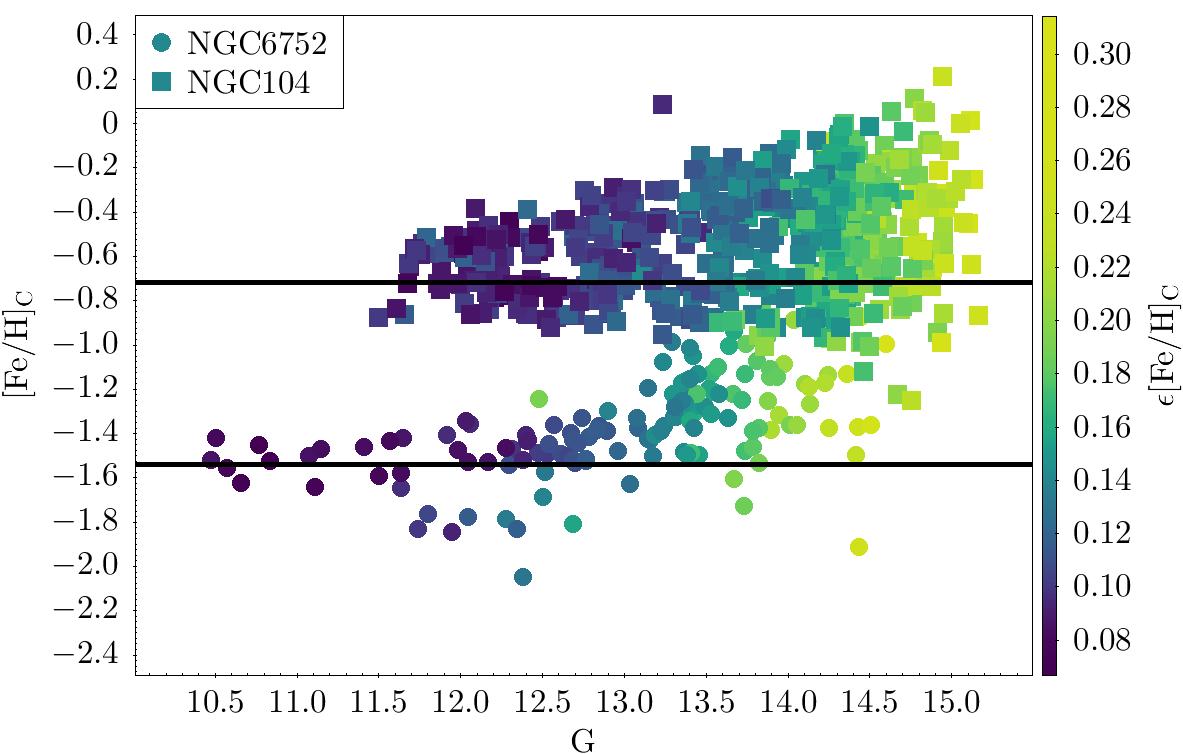}
}
\caption{${\rm [Fe/H]_{C}}$ as a function of apparent G magnitude for member stars of NGC~104 and NGC~6752 included in our catalogue. Points are colour-coded according to their uncertainty in the photometric metallicity $\epsilon{\rm [Fe/H]_{C}}$. The thick horizontal lines mark the value of the mean spectroscopic metallicity as tabulated in the 2010 edition of the \citet{harris1996} catalogue. Note that the red horizontal branch stars of NGC~104 lie around $G\simeq 14.2$.}
\label{fig:gctest}
\end{figure} 
%%%%%%%%%%%%%%%%%%%%%%%%%%%%%%%%%%%%%%%%%

%%%%%%%%%%%%%%%%%%%%%%%%%%%%%%%%%%%%%%%%%%%%%%%%%%%%%%%%%%%%
\begin{table*}[!htbp]
\centering
\caption{\label{tab:gctest} Mean [Fe/H]$_{C}$ for selected globular clusters 
with more than 5 stars in our sample. Comparison with [Fe/H]$_{spec}$}
{
    \begin{tabular}{lccccccc}
NGC  & [Fe/H]$_{spec}$ & lim($\epsilon{\rm [Fe/H]_{C}}$) &$\langle {\rm [Fe/H]_{C}}\rangle$ & err  & $\sigma_{\rm [Fe/H]_{C}}$ & err &  N \\ 
\hline  
104  & -0.72  & 0.40 &-0.54 & 0.01 & 0.18 & 0.01 & 621 \\
104  & -0.72  & 0.15 &-0.57 & 0.01 & 0.17 & 0.01 & 343 \\
104  & -0.72  & 0.10 &-0.62 & 0.02 & 0.13 & 0.01 & 100 \\
6205 & -1.53  & 0.40 &-1.61 & 0.04 & 0.00 & 0.04 &   7 \\
6205 & -1.53  & 0.15 &-1.61 & 0.04 & 0.00 & 0.04 &   6 \\
6218 & -1.37  & 0.40 &-1.38 & 0.03 & 0.13 & 0.03 &  35 \\
6218 & -1.37  & 0.15 &-1.37 & 0.03 & 0.12 & 0.03 &  30 \\
6254 & -1.56  & 0.40 &-1.67 & 0.02 & 0.05 & 0.05 &  32 \\
6254 & -1.56  & 0.15 &-1.68 & 0.03 & 0.06 & 0.04 &  25 \\
6752 & -1.54  & 0.40 &-1.40 & 0.02 & 0.15 & 0.02 & 144 \\
6752 & -1.54  & 0.15 &-1.45 & 0.02 & 0.13 & 0.02 &  87 \\
6752 & -1.54  & 0.10 &-1.51 & 0.02 & 0.07 & 0.02 &  29 \\
6809 & -1.94  & 0.40 &-1.78 & 0.04 & 0.00 & 0.10 &  22 \\
6809 & -1.94  & 0.15 &-1.78 & 0.06 & 0.00 & 0.13 &   5 \\
\hline  
    \end{tabular}
}
\tablefoot{[Fe/H]$_{spec}$ is the spectroscopic iron abundance of the cluster from the 2010 version of the \citet{harris1996} catalogue.
lim($\epsilon{\rm [Fe/H]_{C}}$) is the threshold in $\epsilon{\rm [Fe/H]_{C}}$ imposed to the cluster sample before computing  
the mean metallicity, e.g. lim($\epsilon{\rm [Fe/H]_{C}}$)=0.10 means that only stars with 
$\epsilon{\rm [Fe/H]_{C}}\le 0.10$ are considered. Mean ${\rm [Fe/H]_{C}}$, $\sigma_{\rm [Fe/H]_{C}}$, and their respective 
errors are computed with the same Maximum Likelihood algorithm used in \citet{muccia2012}. N is the dimension of the considered sample.
}
\end{table*}
%%%%%%%%%%%%%%%%%%%%%%%%%%%%%%%%%%%%%%%%%%%%%%%%%%%%%%%%%%%% 

In Fig.~\ref{fig:gctest} we show ${\rm [Fe/H]_{C}}$ as a function of apparent G magnitude for stars in the two clusters with the largest number of members included in our catalogue, NGC~104 (47~Tuc) and NGC~6752. The distributions seems reasonably well behaved: both the scatter about the mean and the uncertainties on individual 
${\rm [Fe/H]_{C}}$ increase with increasing G (fainter magnitudes). A systematic shift occurs for G$>13.0$ in the NGC~6752 sample. This can be attributed to the fact that this cluster is very concentrated \citep[concentration parameter C=2.5,][]{harris1996}, hence crowding may have a significant impact at faint magnitudes. On the other hand, it is interesting to note that many stars belonging to the red HB are present in the NGC~104 sample and they behave indistinguishably from RGB stars in Fig.~\ref{fig:gctest}. We verified that the same holds also for AGB stars, within the uncertainties. The diagram suggests that both G magnitude and $\epsilon{\rm [Fe/H]_{C}}$ can be used to select high-quality sub-samples from our catalogue.

In Table~\ref{tab:gctest} we compare the mean metallicity obtained from GC stars in our catalogue with [Fe/H]$_{spec}$ from \citet{harris1996}. Mean metallicities and standard deviations are derived taking into account the individual uncertainties, by means of the maximum likelihood (ML) algorithm used in \citet{muccia2012}, and originally developed for other purposes by \citet{pm93}. Different cuts on $\epsilon{\rm [Fe/H]_{C}}$ have been adopted, when possible, to explore the effect on the accuracy of the mean ${\rm [Fe/H]_{C}}$ obtained. The main results of this analysis can be summarised as follows: 1. independently of the adopted $\epsilon{\rm [Fe/H]_{C}}$ cut, the derived mean photometric metallicities are within $\la 0.15$ of the spectroscopic metallicity, 2. cuts in $\epsilon{\rm [Fe/H]_{phot}}$ result in more accurate mean metallicities, 3. errors on the mean metallicity are, in some cases, significantly smaller than the difference between [Fe/H]$_{spec}$ and [Fe/H]$_{C}$, suggesting that individual uncertainties are somehow underestimated, as expected, 4. probably for the same reason, in some case 
$\sigma_{[Fe/H]_{C}}$ values not consistent with zero are found.

%%%%%%%%%%%%%%%%%%%%%%%%%%%%%%%%%%%%%%%%%%%%%%%%%%%%%%%%%%%%%%%%%%%%%%%%%%%%
\begin{table}[!htbp]
\centering
\caption{\label{tab:Omed} Mean [Fe/H]$_{phot}$ for selected open clusters 
with more than 5 stars in our sample. Comparison with [Fe/H]$_{spec}$}
{
    \begin{tabular}{lccccc}
NGC  & [Fe/H]$_{spec}$ & $\langle {\rm [Fe/H]_{C}}\rangle$ & $\langle {\rm [Fe/H]_{H}}\rangle$ & N \\
\hline  
188  &  0.07$\pm$ 0.04 & -0.18$\pm$ 0.03 &  0.18$\pm$  0.05 &  23\\
752  & -0.06$\pm$ 0.01 & -0.12$\pm$ 0.05 &  0.09$\pm$  0.05 &   8\\
2682 &  0.00$\pm$ 0.05 & -0.27$\pm$ 0.02 & -0.01$\pm$  0.05 &  15\\
\hline  
    \end{tabular}
}
\tablefoot{[Fe/H]$_{spec}$ is the mean spectroscopic iron abundance from 
\citet{myers22}. The mean photometric metallicities and their
respective  errors are computed with the same Maximum Likelihood algorithm used in
\citet{muccia2012}. N is the dimension of the considered sample. 
}
\end{table}
%%%%%%%%%%%%%%%%%%%%%%%%%%%%%%%%%%%%%%%%%%%%%%%%%%%%%%%%%%%%%%%%%%%%%%%%%%%%

Well populated open clusters (OCs) are even rarer than GCs in our sample, mainly due to the selection imposed on Galactic latitude, reddening, and, implicitly on age, since only clusters older than $\simeq 1-2$~Gyr have RGB stars. In Tab.~\ref{tab:Omed} we show a comparison analogous to that performed for GCs for three OCs having more than five members included in our sample, NGC~752, NGC~188, and NGC~2682 (M~67), whose age is 1.2~Gyr, 7.1~Gyr, and 4.3~Gyr, respectively \citep{myers22}. We compare the mean spectroscopic metallicities recently provided by \citet[][based on APOGEE DR17]{myers22} with the means obtained from both our photometric metallicities.
The absolute differences with the spectroscopic values are in the range 0.06-0.27~dex for ${\rm [Fe/H]_{C}}$, and 0.01-0.11 for ${\rm [Fe/H]_{phot,H}}$, confirming that indeed the latter indicator is more accurate in the metallicity/age regime covered by these clusters.

\section{The impact of interstellar extinction on photometric metallicity}
\label{appe_ebv}

Uncertainty and unaccounted for systematic errors in the extinction and, possibly, in the adopted extinction laws, can be a major limitation for the accuracy and the precision of photometric metallicities, especially when derived from medium-width passbands as in our case. Here we have adopted the values from \citet{sfd98} maps, recalibrated following \citet[][S11]{Shlafly2011}, for simplicity, as they are the most widely used. However these are asymptotic values for large distances and may not be appropriate for the nearest stars. According to \citet{lallement19}, in most intermediate latitude directions the extinction profile stabilises near the asymptotic value for $D\ga 1-2$~kpc. Mainly due to the selection in Galactic latitude, only $\le 14\%$ of the stars in our sample lie at $D\le 1.0$~kpc, and we consider only stars with E(B-V)$<0.3$, hence we do not expect our choice to have a major impact on our overall metallicity scale, while it can make a significant difference on individual stars.

\begin{figure}[ht!]
\center{
\includegraphics[width=\columnwidth]{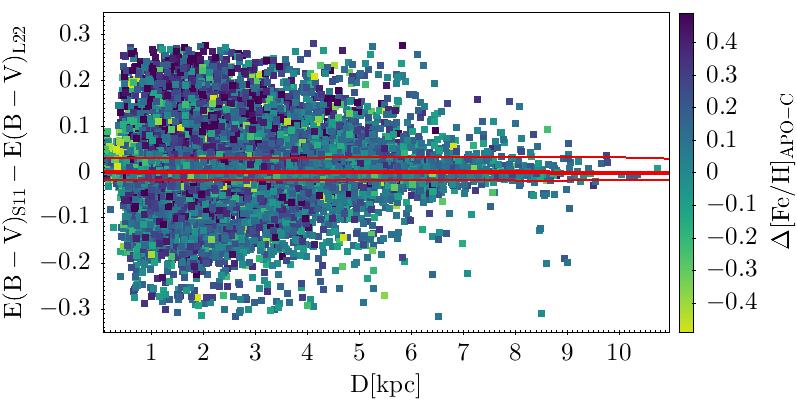}
}
\caption{Difference between the reddening estimates adopted here, from \citet[][S11]{Shlafly2011}, and those obtained from the distance-dependent maps by \citet[][L22]{lallement22} as a function of distance from the Sun.
Points are colour-coded according to ${\rm \Delta [Fe/H]_{APO-C}=[Fe/H]_{APO}-[Fe/H]_{C}}$. The thick red line is $P_{50}$ and the thin red lines are $P_{16}$ and $P_{84}$ of the distribution of the E(B-V) difference computed over 0.5~kpc wide bins.}
\label{fig:ebvdiff}
\end{figure}

In Fig.~\ref{fig:ebvdiff} we plot the difference between the S11 reddening values we adopted and those derived from the distance-dependent maps by \citet[][L22]{lallement22}, as a function of distance for the APOGEE DR17 VS. While the boundaries of the $\Delta E(B-V)$ distribution shrinks with increasing distance, as expected, the (asymmetric) $\pm 1\sigma$ interval remains remarkably constant, and the median difference very close to zero, over the entire range of distances. Hence, the two reddening scales are in good agreement for the bulk of stars in our sample. Moreover, asymptotic $E(B-V)_{S11}$ are expected to systematically overestimate reddening for nearby stars. While there is some asymmetry toward positive values in the $\Delta E(B-V)$ distribution at low distances, with stars deviating by more than +0.05~mag from zero being two times more abundant than those with $\Delta E(B-V)<-0.05$, negative deviations are not so rare and display amplitudes comparable with those observed on the positive side. This strongly suggest that both the considered sources of reddening estimates bring in their own systematic errors. Indeed, at low $E(B-V)_{L22}$ values S11 seems to overestimate the extinction for a fraction of stars, while at high $E(B-V)_{L22}$ the opposite occurs, with the distance-dependent estimates significantly exceeding the asymptotic values from S11. Hence, each choice of the reddening scale comes at a comparable cost in terms of errors in the photometric metallicity.

%%%%%%%%%%%%%%%%%%%
%%%%%%%%%%%%%%%%%%%%%%%%%%%%%%%%%%%%%%%%%

%%%%%%%%%%%%%%%%%%%
\begin{figure}[ht!]
\center{
\includegraphics[width=\columnwidth]{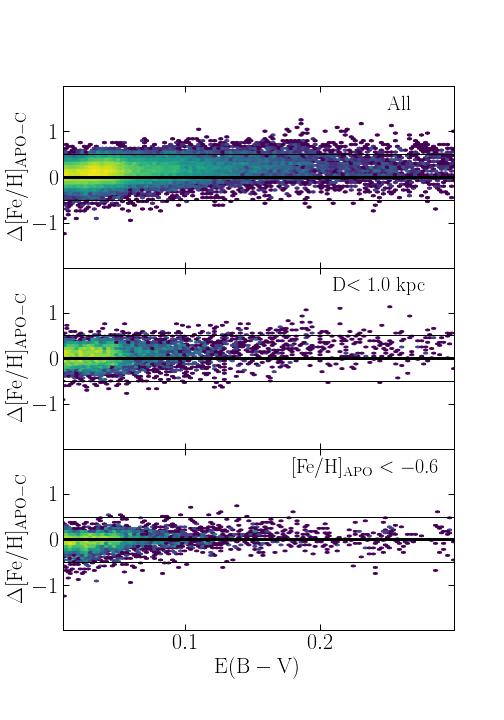}
}
\caption{${\rm \Delta [Fe/H]_{APO-C}}$  as a function of ${\rm E(B-V)}$ for the APOGEE VS. Upper panel: entire sample, middle panel: stars with distance lower than 1~kpc, lower panel: stars with 
${\rm [Fe/H]_{APO}<-0.6}$. The horizontal lines mark the 
${\rm \Delta [Fe/H]}=0.0\pm 0.5$~dex levels.}
\label{fig:ebvtrend}
\end{figure} 
%%%%%%%%%%%%%%%%%%%%%%%%%%%%%%%%%%%%%%%%%

The colour-coding according to ${\rm \Delta [Fe/H]_{APO-C}}$ shows that some broad correlation between $\Delta E(B-V)$ and ${\rm \Delta [Fe/H]_{APO-C}}$ is there but it is very weak (Pearson's correlation coefficient = +0.102), showing that reddening errors are not the main driver of correlated metallicity errors (see below). Analogous conclusions can be drawn if the analysis is repeated for ${\rm[Fe/H]_H}$.

In Fig.~\ref{fig:ebvtrend} we explore possible dependencies of ${\rm [Fe/H]_C}$ from the adopted reddening values. The upper panel of Fig.~\ref{fig:ebvtrend} shows that, indeed, ${\rm \Delta [Fe/H]}$ strongly correlates with ${\rm E(B-V)}$, with an amplitude reaching $\simeq 0.2$~dex for  ${\rm E(B-V)}\ge 0.2$. This does not affect the bulk of the stars in our sample, as only 4\%(21\%) of them have $E(B-V)>0.2(0.1)$.
The sense of the trend is that photometric metallicities under-estimate their spectroscopic counterparts.

However the middle and lower panels of the same figure provide robust evidence that this is a secondary correlation, i.e., a manifestation of the trend of ${\rm \Delta [Fe/H]_{APO-C}}$ with ${\rm [Fe/H]_{spec}}$ arising at ${\rm [Fe/H]_{spec}}\ga -0.5$, discussed in Sect.~\ref{apogee} and App.~\ref{appe_val}. This occurs because more reddened sources are more likely to lie at low Z, and, consequently, have higher metallicity, in the regime were the quoted trend has sizeable amplitude. 

In the middle panel of Fig.~\ref{fig:ebvtrend} we show that the subset of the nearest stars (${\rm D<1.0}$~kpc) displays the same trend as the entire sample. Hence, the observed trend is not driven by nearby stars over-corrected by our assumption of asymptotic reddening values. This result also supports the validity of our choice of the source of reddening values.

The lower panel of the figure shows that stars with
${\rm [Fe/H]_{spec}}\le -0.6$ do not display any appreciable trend of ${\rm \Delta [Fe/H]}$ with ${\rm E(B-V)}$, thus confirming that the effect seen in the uppermost panel is indeed driven by the most metal-rich stars, also providing further support to the reliability of our ${\rm [Fe/H]_{C}}$ estimates in the metallicity regime we are mostly interested in.
We also verified that the trend becomes more and more evident as we include more metal-rich stars. All of these results support the idea that the trend in the upper panel of Fig.~\ref{fig:ebvtrend} is indeed a secondary effect of the trend of ${\rm \Delta [Fe/H]_{APO-C}}$ with ${\rm [Fe/H]_{spec}}$ in the metal-rich regime\footnote{This, in turn, can be the secondary effect of a trend primarily driven by age, as suggested in Appendix~\ref{appe_trend}}. Consistent with this conclusion, the trend of ${\rm \Delta [Fe/H]_{APO-H}}$ with ${\rm [Fe/H]_{spec}}$ has been verified to be  significantly weaker than that of ${\rm \Delta [Fe/H]_{APO-C}}$.

\section{Trends with [$\alpha$/Fe]}
\label{appe_trend}

$\alpha$ elements (like, e.g., O, Mg, Si, Ca, Ti) are among the most abundant metals in the atmosphere of stars and their strong lines may have a non-negligible impact on photometric metallicities, especially as the $[\alpha/{\rm Fe}]$ distribution at a given metallicity is known to be be strongly bimodal in the Galaxy, at least for ${\rm [Fe/H]}\ga -1.5$ \citep[see, e.g.,][and references therein]{hasse21,hayden22}.

%%%%%%%%%%%%%%%%%%%
\begin{figure}[ht!]
\center{
\includegraphics[width=\columnwidth]{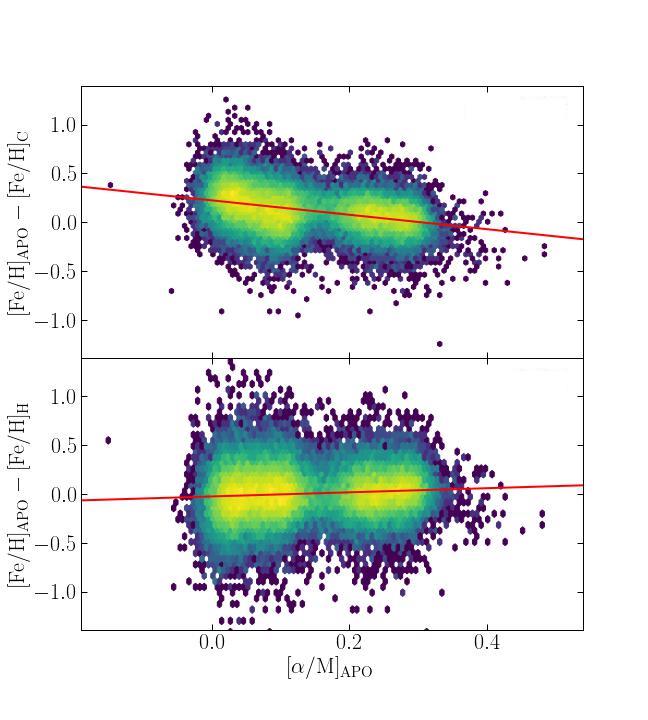}
}
\caption{${\rm \Delta [Fe/H]_{APO-C}}$  (upper panel) and ${\rm \Delta [Fe/H]_{APO-H}}$ (lower panel) as a function of mean spectroscopic $\alpha$ elements abundance for the APOGEE VS. The red lines are linear fits to the data.}
\label{fig:alfadelta}
\end{figure} 
%%%%%%%%%%%%%%%%%%%%%%%%%%%%%%%%%%%%%%%%%

For this reason it may be worth checking if our photometric metallicities display some dependency on $\alpha$ elements abundance. In Fig.~\ref{fig:alfadelta} we show the distribution of ${\rm \Delta [Fe/H]_{APO-C}}$ and ${\rm \Delta [Fe/H]_{APO-H}}$ as a function of the mean $\alpha$ elements abundance provided by APOGEE $[\alpha/M]$, for the APOGEE DR17 VS. Indeed, ${\rm \Delta [Fe/H]_{APO-C}}$ shows a strong anti-correlation with $[\alpha/M]$: ${\rm [Fe/H]_C}$ tend to underestimate the spectroscopic metallicity with decreasing $[\alpha/M]$. Similar results are obtained for all the considered VSs. 

This is the same kind of trend observed for increasing ${\rm [Fe/H]_{spec}}$ for 
${\rm [Fe/H]_{spec}}\ga -0.5$. Since in that metallicity range the sample is dominated by low ${\rm [\alpha/Fe]}$ stars, it seems plausible that the correlation is mainly a by-product of the ${\rm \Delta [Fe/H]_{APO-C}}$ trend with metallicity, as in the case of the correlation with reddening discussed in Appendix~\ref{appe_ebv}. The fact that ${\rm \Delta [Fe/H]_{APO-H}}$ displays a much weaker correlation with $[\alpha/M]$, and in the opposite sense, confirms this hypothesis. However, given the strong correlation between age and ${\rm [\alpha/Fe]}$ in the solar neighbourhood \citep[see, e.g.][]{haywood13,bensby14,hayden17,hayden22}, it cannot be excluded that differences in age are also playing a role, in the high-metallicity regime (see Sect.~\ref{cal}). In fact, it is possible that the onset of the dependency on age at high metallicity (Sect.~\ref{cal}) is the original driver of the trend with metallicity of ${\rm [Fe/H]_C}$, in that metallicity range.

\end{appendix}

\end{document}